%% file: main.tex
\PassOptionsToPackage{hyphens}{url}
\documentclass[sigconf,natbib=false,screen,10pt]{acmart}

\usepackage[utf8]{inputenc}
\usepackage[T1]{fontenc}
\usepackage[american]{babel}
\usepackage{csquotes}

\usepackage{todonotes}
\usepackage{listings}
\usepackage{graphicx}

\usepackage{float}
\newfloat{algorithm}{tb}{lop}

\usepackage{stfloats}
\usepackage{graphicx}
\usepackage{caption}
\usepackage{subcaption}
\usepackage{xspace}
\usepackage{amsmath}
\usepackage{amssymb}
\usepackage{booktabs}
\usepackage{dcolumn}
\usepackage{tabularx}
\usepackage{mathtools}
\usepackage[binary-units]{siunitx}
\usepackage{pifont}

\usepackage{tikz}
\usetikzlibrary{calc}
\usetikzlibrary{fit}
\usetikzlibrary{positioning}
\usetikzlibrary{shapes.symbols}

\tikzset{%
  every fit/.style={transform shape=false},
}

\DeclareSIUnit{\nothing}{\relax}
\DeclareSIUnit{\x}{\times}
\DeclareSIUnit\invocations{\text{invocations}}
\newcommand{\factor}[1]{\SI{#1}{\x}}

\usepackage{algorithm}
\usepackage[noend]{algpseudocode}
\algrenewcommand\algorithmicfunction{\textbf{func}}
\algrenewcommand\algorithmicprocedure{\textbf{program}}
\colorlet{AlgCommentGray}{black!80!white}
\algrenewcommand\algorithmicindent{1.5em}
\makeatletter
\renewcommand{\ALG@beginalgorithmic}{\small}
\makeatother

\useshorthands*{"}
\defineshorthand{"-}{\babelhyphen{hard}}

\PassOptionsToPackage{
    sortcites,
    maxcitenames=2,
    maxbibnames=8,
  }{biblatex}
\usepackage{biblatex}
\bibliography{main}

\usepackage{xcolor}
\definecolor{ETHa}{RGB}{31,64,122}      
\definecolor{ETHb}{RGB}{72,90,44}       
\definecolor{ETHc}{RGB}{18,105,176}     
\definecolor{ETHd}{RGB}{114,121,28}     
\definecolor{ETHe}{RGB}{145,5,106}      
\definecolor{ETHf}{RGB}{111,111,100}    
\definecolor{ETHg}{RGB}{168,50,45}      
\definecolor{ETHh}{RGB}{0,122,150}      
\definecolor{ETHi}{RGB}{149,96,19}      

\PassOptionsToPackage{pdfpagelabels=false}{hyperref}
\usepackage{hyperref}

\hypersetup{%
    colorlinks=true,%
    urlcolor=ETHg,%
    linkcolor=ETHc,%
    citecolor=ETHd,%
    breaklinks=true,%
    pageanchor=true,%
}

\makeatletter
\newcommand{\anonymize}[2]{%
  \if@ACM@anonymous#1\else#2\fi\xspace}
\makeatother

\newcommand{\bigquery}{Google BigQuery\xspace}
\newcommand{\athena}{Amazon Athena\xspace}

\newcommand{\qaas}{QaaS\xspace}
\newcommand{\faas}{FaaS\xspace}
\newcommand{\iaas}{IaaS\xspace}
\newcommand{\paas}{PaaS\xspace}
\newcommand{\saas}{SaaS\xspace}
\newcommand{\Qaasl}{Query-as-a-Service\xspace}

\newcommand{\lambada}{{\texttt Lambada}\xspace}

\makeatletter
\DeclareRobustCommand{\varname}[1]{\begingroup\newmcodes@\mathit{#1}\endgroup}

\newcommand{\sslash}{\mathbin{/\mkern-6mu/}}

\newcommand{\Oh}{\mathcal{O}}

\definecolor{darkgreen}{rgb}{0.18,0.54,0.34}
\definecolor{codegreen}{rgb}{0,0.6,0}

\lstdefinelanguage{CVMPY}[]{Python}{%
    morekeywords=[2]{from_parquet,filter,map,reduce},
}

\lstdefinestyle{cvmpy} {
    language=CVMPY,
    keywordstyle=\itshape,
    keywordstyle=[2]\color{orange!50!black},
    stringstyle=\color{red!50!brown},
}

\setcopyright{iw3c2w3}

\begin{document} 
\title[Lambada: Interactive Data Analytics on Cold Data using Serverless Cloud Infrastructure]
      {Lambada: Interactive Data Analytics \\ on Cold Data using Serverless Cloud Infrastructure}

\author{Ingo Müller}
\email{ingo.mueller@inf.ethz.ch}
\orcid{0000-0001-8818-8324}
\affiliation{}

\author{Renato Marroquín}
\email{marenato@inf.ethz.ch}
\affiliation{}

\author{Gustavo Alonso}
\email{alonso@inf.ethz.ch}
\affiliation{}

\renewcommand{\shortauthors}{I. Müller, R. Marroquín, G. Alonso}

\copyrightyear{2020}
\acmYear{2020}
\acmConference[SIGMOD'20]{ACM SIGMOD International Conference on Management of Data}{June 14--19, 2020}{Portland, OR, USA}
\acmBooktitle{SIGMOD'20}
\acmPrice{15.00}
\acmDOI{12.345/67.89}
\acmISBN{978-1-2345-6789-1/18/10}

\begin{abstract}
  The promise of ultimate elasticity and operational simplicity
  of serverless computing
  has recently lead to an explosion of research in this area.
  In the context of data analytics, the concept sounds appealing,
  but due to the limitations of current offerings,
  there is no consensus yet on
  whether or not this approach is technically and economically viable.
  In this paper, we identify interactive data analytics on cold data
  as a use case where serverless computing excels.
  We design and implement \emph{Lambada},
  a system following a purely serverless architecture,
  in order to illustrate when and how serverless computing
  should be employed for data analytics.
  We propose several system components
  that overcome the previously known limitations
  inherent in the serverless paradigm
  as well as additional ones we identify in this work.
  We can show that, thanks to careful design,
  a serverless query processing system
  can be at the same time one order of magnitude faster
  and two orders of magnitude cheaper
  compared to commercial \Qaasl systems,
  the only alternative with similar operational simplicity.
\end{abstract}

\maketitle 

\input{introduction}
\input{background}
\input{architecture}
\input{serverless_operators}
\input{experiments}
\input{relwork}
\input{conclusions}

\printbibliography

\end{document}

%% file: introduction.tex
\section{Introduction}

Data processing in the cloud has become a widespread solution
in a wide variety of use cases.
Over time, deployment models and levels of abstraction
have undergone a tremendous evolution.
While in the early days of Infrastructure-as-a-Service (\iaas),
the cloud mainly consisted in providing bare computing resources
in the form of virtual machines (VMs),
it offers now a richer computing and development experience
through Platform-as-a-Service (\paas).
In both cases, the basic assumption is that the cloud is used
as a rented computing infrastructure,
whose elasticity can lead to a lower total cost than owned infrastructure.
However, elasticity is limited by
how fast the infrastructure can be started and stopped, and services migrated.
Thus, cloud offerings evolved further
towards Software-as-a-Service (\saas) in a variety of forms,
where customers do not rent infrastructure \emph{per se}
but the use of a given software functionality.
In data processing, such systems have been quite successful:
\bigquery~\cite{big_query},
\athena~\cite{athena}, etc.
are examples of \Qaasl (\qaas) systems
that provide database services without actually having to run (and pay for)
a database engine as part of the cloud usage.

The demand for even higher elasticity and more fine-grained billing
has recently led to the proliferation of Function-as-a-Service (\faas) systems.
\faas is a way to implement serverless computing---%
a name that emphasizes precisely the advantages of the approach:
there is no need to install, operate, and manage a server (infrastructure)
to get computations done.
This has lead to successful applications such as
source code compilation~\cite{Klimovic2018b,fouladi19} or
video encoding~\cite{fouladi19,sprocket18}.

\begin{figure}
    \centering
    \begin{subfigure}[t]{.475\linewidth}
        \centering
        \includegraphics[width=.9\linewidth,height=3.8cm]{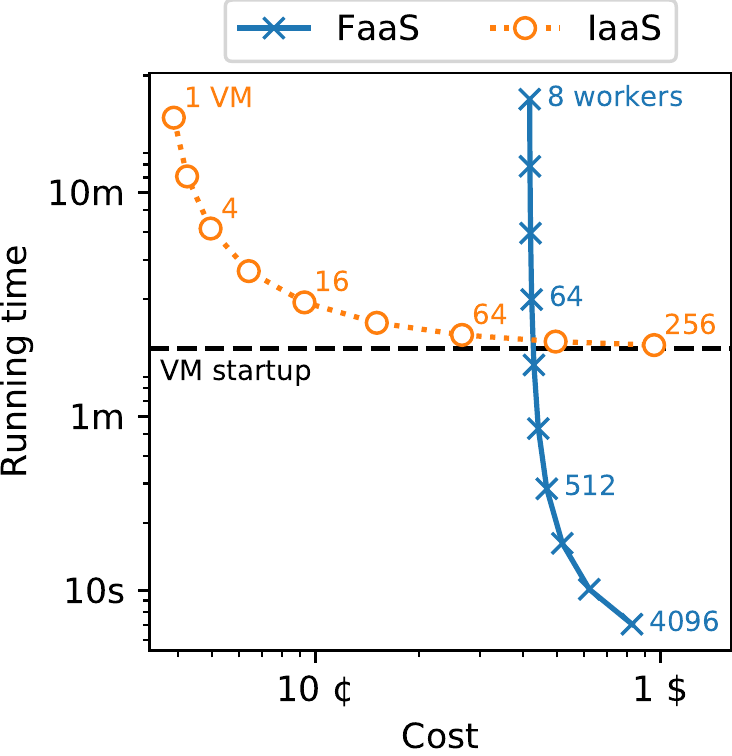}
        \caption{Job-scoped resources.}
        \label{fig:intro-job-scoped-resources}
    \end{subfigure}
    \hfill
    \begin{subfigure}[t]{.475\linewidth}
        \centering
        \includegraphics[width=\linewidth]{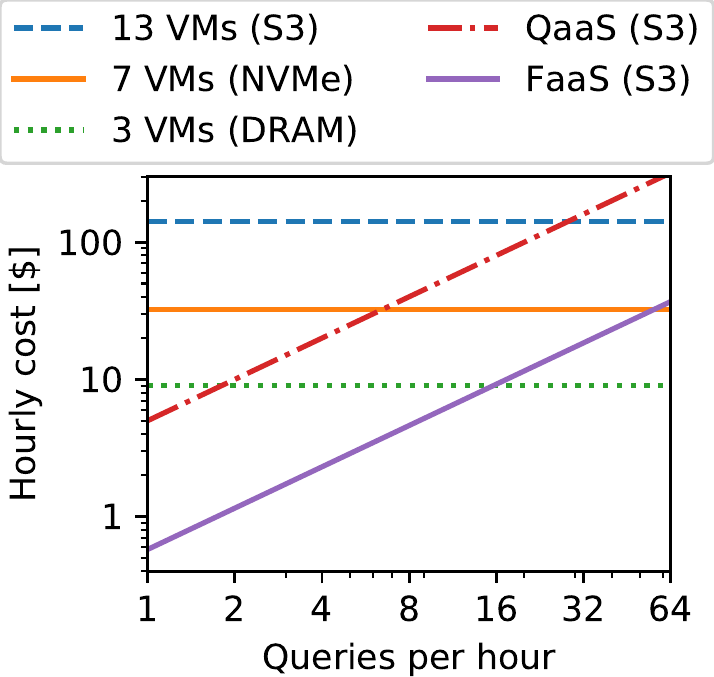}
        \caption{Always-on resources.}
        \label{fig:intro-always-on-resources}
    \end{subfigure}
    \caption{Comparison of cloud architectures.}
    \label{fig:intro-serverless-characteristics}
    \Description{Comparison of cloud architectures.}
\end{figure}

In order to understand the types of applications
where \faas is particularly attractive in the context of data analytics,
consider a query scanning \SI{1}{\tera\byte} of data stored
on Amazon Simple Storage Service (S3).
There are two ways to use \iaas for this task:
starting a set of resources for the duration of a single job
(``job-scoped'' resources)
or scheduling jobs onto resources that are kept running
(``always-on'' resources).
Figure~\ref{fig:intro-job-scoped-resources} shows the costs and running time
of job-scoped resources obtained through simulation
for a varying number of workers.%
\footnote{Between 1 and 256 \texttt{c5n.xlarge} instances
          and between 8 and 4096 concurrent function invocations
          with \SI{2}{\gibi\byte} main memory, respectively.}
As the plot shows, in both cases,
adding more resources reduces the running time,
but with a diminishing gain as we approach the respective startup time.%
\footnote{We assume \SI{2}{\minute} start-up time for \iaas
          and \SI{4}{\second} for \faas.}
To obtain the lowest cost, \iaas is thus more attractive,
being up to an order of magnitude cheaper.
However, if query latency is important,
even if that means a higher price,
then \faas is more attractive.
The strength of \faas compared to job-scoped \iaas
in data analytics is thus the ability to service \emph{interactive} queries.

An alternative way to use \iaas is to keep resources running.
To understand the trade-offs of that approach compared to \faas,
consider the case of a data processing system
running permanently in enough VMs
such that the query above can be processed in under \SI{10}{\second}.
This simulates an application that cannot be solved with job-scoped \iaas
due to the start-up overhead.
In this scenario, the system can load the data up-front
into different levels of the memory hierarchy
and scan it at the bandwidth of the respective level.
In order to meet the \SI{10}{\second} latency target,
this requires three large instances
if we load and read the data from fast DRAM,
seven of the largest instances if using somewhat slower NVMe,
and thirteen instances
if we process the data directly from S3 without pre-loading.%
\footnote{Our simulation uses \texttt{r5.12xlarge}, \texttt{i3.16xlarge},
          and \texttt{c5n.18xlarge} instances, respectively.}
Figure~\ref{fig:intro-always-on-resources} shows the expected hourly cost
of the different configurations as a function of the query frequency.
Running virtual machines incurs only hourly costs,
which is independent of the frequency at which queries are run.
In contrast, \faas and \qaas have a usage-based pricing model.
Price increases linearly with the number of queries,
such that even a moderate query load makes them more expensive than \iaas.
The strength of \faas compared to always-on \iaas
is thus for \emph{sporadic use}.
In other words, the stellar use case of data analytics on \faas
is that of the ``lone-wolf data scientist,''
who runs a small number of interactive queries
on datasets only she is working with.

However, using \faas as compute resources is controversial
since serverless functions come with significant limitations:
restricted network connectivity, limited running time,
essentially stateless operation with a very limited cache between invocations,
and lack of control over the scheduling of functions.
All of these shortcomings have been recently analyzed
and clearly spelled out in the literature%
~\cite{hell_lambdas,Klimovic2018,Klimovic2018b,pywren}.

In the context of data analytics,
the most severe limitation of \faas is arguably
the inability to accept network connections from the outside,
rendering direct communication between function invocations impossible.
Previous work proposes a number of approaches,
all of which consist in running some kind of service
in additional infrastructure running on traditional VMs,
either with existing object caches such as Redis~\cite{pywren}
or a service for ephemeral storage
tailored to serverless computing~\cite{Klimovic2018,Pu2019}.
These approaches are technically interesting
and do enable the functions to communicate
but at the cost of reintroducing always-on infrastructure.
Any such ``serverful'' component has the potential
to severely limit the attractiveness of \faas---%
as shown by the introductory example,
it either sacrifices interactivity
(in which case the whole query could run in VMs at a much lower price)
or it introduces a constant cost per time,
which can dominate the cost at the infrequent use that \faas shines at.

Motivated by the successful usage of \faas in other domains,
this paper addresses the question of whether \faas
can be efficiently and effectively used for data analytics.
With the observations drawn from the simulations above,
\faas is most attractive
for \emph{interactive data analytics on cold data}.
Specifically, we (1) identify the technical limitations
of a concrete implementation of \faas, AWS Lambda;
(2) propose suitable solutions to the limitations
that do not fundamentally reduce their economic advantage,
i.e., solutions that require \emph{only serverless components};
and (3) clarify the use cases
in which the cost model behind Lambdas makes sense.
In particular, we design a number of data processing components
that accommodate
the existing limitations of serverless cloud infrastructure
(known ones and others we identify in the paper)
to build \emph{Lambada}, a scalable data processing system,
which does not rely on any ``always-on'' infrastructure.
Lambada is able to answer queries
over gigabytes to terabytes of cold data at interactive query latency.
In the most favorable cases,
it is at the same time two orders of magnitude cheaper
and one order of magnitude faster than commercial \qaas offerings.

The paper makes the following contributions:
\begin{itemize}
  \item We characterize interactive analytics on cold data
    as the sweet spot for using \faas.
  \item We show that AWS Lambda currently
    exposes a small amount of intra-function parallelism
    and how to exploit it.
  \item We identify the process of invoking a large number of functions naively
    as incompatible with the interactivity requirement
    and propose an approach with sublinear runtime
    that can spawn 4k functions in \SI{3}{\second}.
  \item We describe the effect of the input block size
    on the performance and monetary cost of reading data from cloud storage
    and design an efficient scan operator
    that fully exploits the available network bandwidth.
  \item We implement these components in a data processing system
    and compare its performance to other serverless solutions,
    characterizing the competitiveness of \faas in this domain.
  \item We design a purely serverless exchange operator
    that overcomes the rate limit of cloud storage faced by previous works
    by reducing the request complexity to a sub-quadratic amount.
\end{itemize}

%% file: background.tex
\section{Background}

\subsection{Usage Model of \lambada}
\label{subsec:background:lambada_usage_model}

\begin{figure}
  \centering
  \includegraphics[width=.8\linewidth]{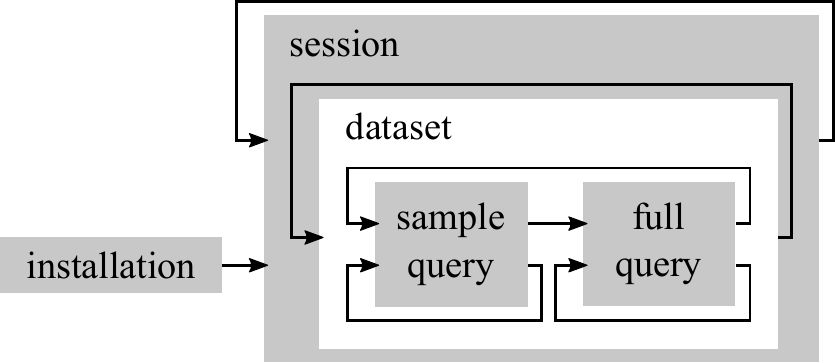}
  \caption{Usage model of data analytics systems.}
  \label{fig:usage-model}
\end{figure}

In order to fix the terminology
for describing how users interact with data analytics systems,
we define a model of how a single user uses such a system.
Figure~\ref{fig:usage-model} illustrates that model:
At some point, the user \emph{installs} the system.
Subsequently, they work with it during short \emph{sessions}.
During each session, the user might work with several \emph{datasets},
each of which he/she might first explore by working on a \emph{sample}
before moving on to run the query on the \emph{full} dataset.

The work of the user is thus characterized by the queries themselves,
but also by the time between them.
There is a break between sessions,
typically in the order of hours to weeks,
and (user) think time between queries,
typically in the orders of seconds to minutes.
The think time includes activities such as analyzing and visualizing the results,
writing and debugging query code,
and making plans about what to do next.

\subsection{Definition and Examples of Serverless Computing}

For the purpose of this paper,
we define the notion of \emph{serverless computing}
in the context of data analytics systems
based on the usage model described above.
Specifically, we define \emph{serverless} to denominate systems or components
that \emph{only incur costs for executing queries} (sample or full).
In particular, serverless systems do not incur costs
for installation, idle infrastructure
during think time or between sessions, or switching datasets.

Cloud providers offer a number of services at various levels of abstraction
that qualify for this definition of \emph{serverless}.
At the highest level of abstraction are interactive query systems,
sometimes called \Qaasl (\qaas),
where the user pays only for each individual query they run.
Examples include \athena~\cite{athena}
and \bigquery~\cite{big_query},
which we evaluate as alternatives to our system
in Section~\ref{sec:experiments}.

Furthermore, there is a number of lower-level serverless services
that can be used to build higher-level ones.
For compute, Amazon offers AWS Lambda, AWS Fargate, and Amazon EC2
to run code in a function, a container, and a virtual machine, respectively.
All of these could be used on a per-query basis
and could thus qualify as \emph{serverless},
but, as we study in more detail below,
only AWS Lambda has low enough start-up times for interactive analytics.
For storage, Amazon offers DynamoDB and S3,
which are both \emph{serverless} in the sense defined above
if used for temporary data during query execution.
Finally, Amazon offers a message queue service (Amazon SQS)
and a workflow service (AWS Step Functions),
whose pure pay-per-use pricing model makes them serverless as well.
Similar services can be found
in the offerings of the other major cloud
providers~\cite{google_cloud_functions,azure_functions}.

In contrast, running a virtual machine or a cluster thereof
for the time of a \emph{session}, no matter what software it runs,
is not serverless by our definition.
The same is true for higher-level cloud services
whose cost depend on the virtual machines they run on.
Examples include Amazon Redshift, Aurora, RDS, and its other managed DBMSs,
Amazon EMR (Elastic MapReduce), and Amazon ElastiCache,
as well as the corresponding offerings from the other providers.

%% file: architecture.tex
\section{Architecture of \lambada}
\label{sec:architecture}

\lambada addresses the question of how to build a \emph{serverless} system for
interactive data analytics using \textit{solely} existing serverless components and how
it compares to \iaas and \Qaasl in terms of performance and price. In the
following sections, we describe its overall architecture and its serverless
components.

\subsection{Overview}

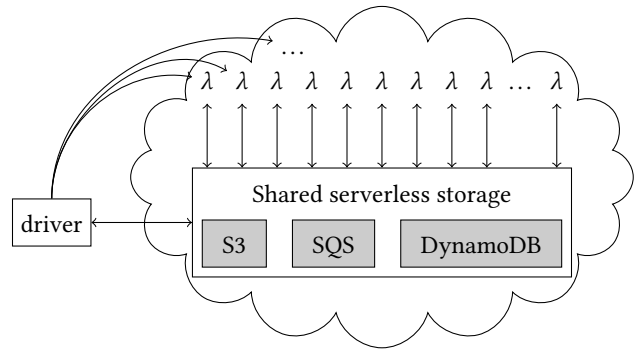
\begin{figure}
  \centering
  \input{figures/architecture-overview}
  \vspace{-1ex}
  \caption{Architecture overview of \lambada.}
  \label{fig:architecture-overview}
\end{figure}

Figure~\ref{fig:architecture-overview} depicts the high-level architecture of
\lambada.
The \emph{driver} runs on the local development machine of the data scientist.
When she executes a query,
the driver invokes a (potentially large) number of \emph{serverless workers}
(depicted as $\lambda$ in the plot),
who execute the query in a data-parallel manner.
The workers communicate through different types
of \emph{shared serverless storage}:
the cloud storage system \emph{AWS S3}
for large amounts of data,
the key-value store \emph{AWS DynamoDB} for small amounts of data,
and the message service \emph{AWS SQS} (Simple Queuing System) for short messages.
Input and output are read from and written to shared storage as well.
In a way, this is a classical shared storage database architecture,
except that \emph{all} communication of the workers
goes through shared storage
and there is no direct communication between the workers or with the driver.
The driver also uses the shared storage
to communicate with the workers once they have been invoked,
for example, to collect the results of their query fragments.

\subsection{Query Compilation}

\begin{lstlisting}[language=python,
    caption={Example query in Python frontend.},
    label=lst:example-query,
    style=cvmpy,
  ]
data = lambada \
    .from_parquet('s3://bucket/*.parquet')
    .filter(lambda x: x[1] >= 0.05)
    .map(lambda x: x[1] * x[2])
    .reduce(lambda x, y: x + y)
\end{lstlisting}

Before a query is executed, it goes through a series of translations
that transform it into an executable form.
We use a query compilation and execution framework
that we are building in our group as part of a larger effort
aiming to run any type of data analytics on a large number of compute platforms
(such as x86 processors, accelerators, or cloud infrastructure).
The framework supports a number of frontend languages,
such as (a subset of) SQL and a UDF-based library interface written in Python,
and is designed to support more frontends in the future.
Listing~\ref{lst:example-query} shows the Python interface of our framework.

We translate the queries of all frontend languages
into a common intermediate representation of query plans,
to which we apply, potentially after some frontend-specific normalizations,
a common set of optimizations such as selection and projection push-downs,
join ordering, and transformation into data-parallel plans.
As a final transformation, we lower pipelines of operators
between materialization points into LLVM IR
and just-in-time compile them to tight native machine code.
For the Python frontend, we use Numba
to translate the UDFs into LLVM IR
and inline them into the remainder of the pipeline,
thus eliminating any interpretation or function calls in the inner loops.

A query plan in our framework is divided into \emph{scopes},
each of which may run in a different target platform.
We currently assign operators to specific scopes
using a combination of annotations and heuristics,
but we believe that cost-based decisions could be added easily in the future.
For instance, most operators in a typical plan of \lambada
run in a serverless scope,
i.e., are executed by the serverless workers.
However, queries may also contain small scopes running on the driver,
in order to do some pre-processing
such as reading small amounts of data locally
that should be broadcasted into the serverless workers
or post-processing like aggregating the intermediate worker results.

\vspace{2ex}
\subsection{Serverless Workers}

The serverless workers run as a \emph{function} in AWS Lambda,
which are set up at installation time.
Such a function consists of an event handler in one of the supported languages,%
\footnote{As of writing, AWS Lambda supports Node.js, Python,
          Java, Ruby, C\#, Go, and PowerShell.}
a ``dependency layer'' that may contain arbitrary native machine code,
and some meta data such as the desired amount of main memory
and the timeout of the function.
The function of serverless workers consists
of a dependency layer containing the same execution framework
that also runs on the driver
and an event handler as wrapper around it implemented in Python.
This event handler extracts the ID of the worker,
the query plan fragment, and its input
from the invocation parameters of the function
and forwards them to the execution framework.
It starts the execution engine in a new process
with a memory limit slightly lower than that of the serverless function
such that it can report out-of-memory situations
and other errors of the execution engine to the driver
rather than dying silently.
When the execution engine finishes its computation,
the handler forwards its results to the driver.
In both cases, if an error occurred or the computation finished successfully,
the handler posts a corresponding message into a \emph{result queue} in SQS,
from which the driver polls until it has heard back from all workers.

%% file: figures/architecture-overview.tex
\newlength\mctextheightarchitecture
\setlength\mctextheightarchitecture{\heightof{$\vec{C}\vec{S}$}}
\begin{tikzpicture}[
      scale=.95,transform shape,
      every node/.style={
        text height=\mctextheightarchitecture,text depth=depth("y$a_0$"),
        node distance=3.5em,},
      storagetype/.style={
        draw,fill=white!80!black,inner xsep=.8em,node distance=1em},
    ]
  \node [storagetype]
    (s3) {S3};
  \node [storagetype,right=of s3]
    (sqs) {SQS};
  \node [storagetype,right=of sqs]
    (dynamodb) {DynamoDB};
  \node [fit=(s3)(sqs)(dynamodb),inner sep=0pt]
    (storage-inner) {};
  \node [above=1ex of storage-inner, inner sep=0pt]
    (storage-label) {Shared serverless storage};
  \node [draw,fit=(storage-inner)(storage-label)]
    (storage) {};

  \node [above=of storage.north west,anchor=west]
    (lambda-1) {$\lambda$};
  \node [above=of storage.north east,anchor=east]
    (lambda-n) {$\lambda$};
  \node (lambda-2)    at ($(lambda-1.center)!0.1!(lambda-n.center)$) {$\lambda$};
  \node (lambda-3)    at ($(lambda-1.center)!0.2!(lambda-n.center)$) {$\lambda$};
  \node (lambda-4)    at ($(lambda-1.center)!0.3!(lambda-n.center)$) {$\lambda$};
  \node (lambda-5)    at ($(lambda-1.center)!0.4!(lambda-n.center)$) {$\lambda$};
  \node (lambda-6)    at ($(lambda-1.center)!0.5!(lambda-n.center)$) {$\lambda$};
  \node (lambda-7)    at ($(lambda-1.center)!0.6!(lambda-n.center)$) {$\lambda$};
  \node (lambda-8)    at ($(lambda-1.center)!0.7!(lambda-n.center)$) {$\lambda$};
  \node (lambda-9)    at ($(lambda-1.center)!0.8!(lambda-n.center)$) {$\lambda$};
  \node (lambda-dots) at ($(lambda-1.center)!0.9!(lambda-n.center)$) {\ldots};

  \draw [<->] (lambda-1.south) -- (lambda-1.south |- storage.north);
  \draw [<->] (lambda-2.south) -- (lambda-2.south |- storage.north);
  \draw [<->] (lambda-3.south) -- (lambda-3.south |- storage.north);
  \draw [<->] (lambda-4.south) -- (lambda-4.south |- storage.north);
  \draw [<->] (lambda-5.south) -- (lambda-5.south |- storage.north);
  \draw [<->] (lambda-6.south) -- (lambda-6.south |- storage.north);
  \draw [<->] (lambda-7.south) -- (lambda-7.south |- storage.north);
  \draw [<->] (lambda-8.south) -- (lambda-8.south |- storage.north);
  \draw [<->] (lambda-9.south) -- (lambda-9.south |- storage.north);
  \draw [<->] (lambda-n.south) -- (lambda-n.south |- storage.north);

  \node[draw,cloud,cloud puffs=16,aspect=1.7,fit=(lambda-1)(lambda-3)(storage),
        inner sep=-.9em,yshift=-.4em]
    (cloud) {};

  \node [draw,left=4em of storage]
    (driver) {driver};
  \coordinate (lambda-3-4) at ($(lambda-3)!.5!(lambda-4)$);
  \node (lambda-invoke-dots) [above=1ex of lambda-3-4] {\ldots};
  \draw [<->] (driver) -- (storage);
  \draw [->]  (driver) to [out=90,in=170] (lambda-1);
  \draw [->]  (driver) to [out=90,in=150] (lambda-2);
  \draw [->]  (driver) to [out=90,in=170] (lambda-invoke-dots);
\end{tikzpicture}

%% file: serverless_operators.tex
\section{System Components for Serverless Analytics}
\label{sec:serverless_ops}

While \lambada's architecture is very similar
to a traditional shared storage database architecture,
implementing such an architecture
in a purely serverless environment is challenging.
In addition to the ones pointed out in previous work,
we identify a number of additional issues
and design system components that overcome all of them.
Each component needs to trade off three things:
(1) hard quotas and limits
from the service-level agreements (SLAs) of the cloud provider
such as a limit on the request rate to S3,
(2) execution speed under the given constraints
(from service limits or from \emph{de facto} performance of a resource),
for example, by overlapping communication with computation,
and (3) usage-based cost of the various serverless services,
such those from the running time of the serverless workers
but also from the number of requests to the various systems.

\subsection{Intra-worker Parallelism}
\label{subsec:intraworker_parallelism}

\begin{figure}
  \centering
  \includegraphics[width=.7\linewidth]{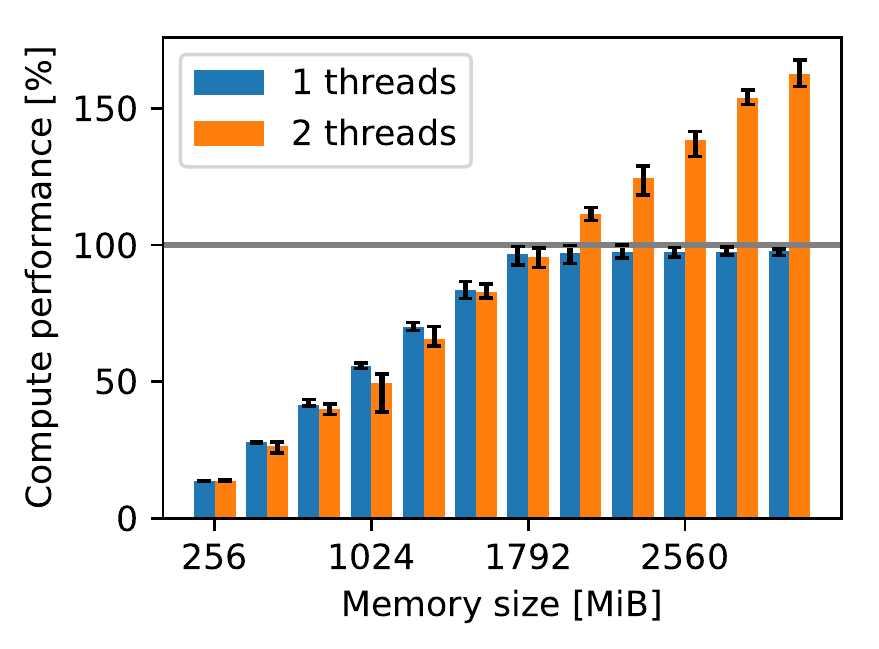}
  \caption{Relative compute performance compared to 1 vCPU.}
  \label{fig:compute-throughput}
\end{figure}

We start with a general observation affecting all subsequent system components:
The functions in AWS Lambda have a small amount of thread-level parallelism.
First, any function can create a relatively large number of threads
(currently up to 1024 as per the service limits%
\footnote{\url{https://docs.aws.amazon.com/lambda/latest/dg/limits.html}}),
which we exploit in various components to overlap network communication.
Second, depending on the function size,
threads may execute on several cores concurrently:
According to the documentation,%
\footnote{\url{https://docs.aws.amazon.com/lambda/latest/dg/resource-model.html}}
the cloud provider allocates an amount of CPU resources to each function
that is proportional to its memory size.
This has been confirmed experimentally by \textcite{Wang2018}.
More precisely, the allocation is such that
a function with \SI{1792}{\mebi\byte} gets the equivalent of one vCPU
and functions with more memory get proportionally more.

We corroborate this statement with a small microbenchmark
shown in Figure~\ref{fig:compute-throughput}.
We run a fixed amount of number crunching operations in functions of various sizes
using either one or two threads
and measure the running time of the computations inside the workers,
i.e., without invocation delay or other overhead.
The functions with \SI{1792}{\mebi\byte} of main memory (using exactly one CPU)
need about \SI{1}{\second}---%
enough to dominate potential overhead of thread scheduling by the OS---%
and set the fastest observed running time in that configuration as baseline.
All measurements are plotted as relative throughput compared to this value.
Indeed, for functions with less than \SI{1792}{\mebi\byte} of main memory,
the compute performance is proportionally lower than the baseline
no matter the number of threads.
Using a single thread, that is the best performance one can achieve,
even with larger functions.
Using a second thread, however, yields proportionally more throughput
with a maximum of \SI{1.67}{\x} the performance of the baseline
for the largest workers with \SI{3008}{\mebi\byte},
indicating that the threads run on more than one CPU.

This observation has the following implication on the system design:
The degree of parallelism is too low
to efficiently exploit data parallelism within a worker,
but may be useful for inter- and intra-pipeline parallelism.
We discuss several opportunities in the remainder of this section.

\subsection{Invocation}
\label{sec:invocation}

\begin{table}
  \centering
  \newcolumntype{N}{D..{3.0}}
  \newcommand{\region}[1]{\multicolumn{1}{c}{\textbf{#1}}}
  \begin{tabular}{@{}lNNNN@{}}
    \toprule
    \textbf{Metric}               & \multicolumn{4}{c}{\textbf{Region}}                       \\
    \cmidrule(lr){2-5}
                                  & \region{eu}  & \region{us} & \region{sa} & \region{ap}    \\
    \midrule
    Single invocation time [ms]   & 36           & 363         & 474         & 536            \\
    Concurrent inv. rate [inv./s] & 294          & 276         & 243         & 222            \\
    Intra-region rate [inv./s]    & 81           & 79          & 84          & 81             \\
    \bottomrule
  \end{tabular}
  \smallskip
  \caption{Characteristics of function invocations.}
  \label{tbl:invocation-characteristics}
\end{table}

As a first component, we discuss how to invoke the serverless workers.
Invoking a large number of them
within a short amount of time is challenging.
Table~\ref{tbl:invocation-characteristics}
shows invocation characteristics of AWS Lambda functions in different regions
from our location in \anonymize{central Europe}{Zurich, Switzerland}.
A single invocation
takes between about \SI{36}{\milli\second} and \SI{0.5}{\second},
depending on the data center and our distance to it.
By overlapping enough concurrent requests at the same time,
we can largely hide the latency of the network round-trip:
By using 128 threads to do the invocations,
we achieve a rate of \SIrange{220}{290}{\invocations/\second}
for any of the data centers.
However, this means that invoking \SI{1000}{\text{workers}}
from the driver still takes \SIrange{3.4}{4.4}{\second}
and linearly more for more workers.
With this approach, the invocation of the serverless workers can thus dominate
the running time of the actual query.%
\footnote{If the query contains a synchronization point
          such that the workers need to wait for each other,
          then this also adds a quadratic component to the monetary cost.}

To reduce the time until all serverless workers are invoked,
we parallelize the invocation process
by off-loading it partially to the first workers.
More precisely, the workers that are invoked by the driver
receive as additional parameter a list of IDs and input data.
Before running their query fragment,
each of this first generation of workers
invokes a second-generation worker for each ID/input pair in that list.
As serverless workers can invoke other workers
at a rate that is in the same ballpark as that of the driver
(see Table~\ref{tbl:invocation-characteristics}),
a reasonable approach is to assign the same amount of invocations
to the driver and to each of the first-level workers,
i.e., about $\sqrt{P}$ invocations each,
where $P$ is the total number of workers.

\begin{figure}
  \centering
  \includegraphics[width=.7\linewidth]{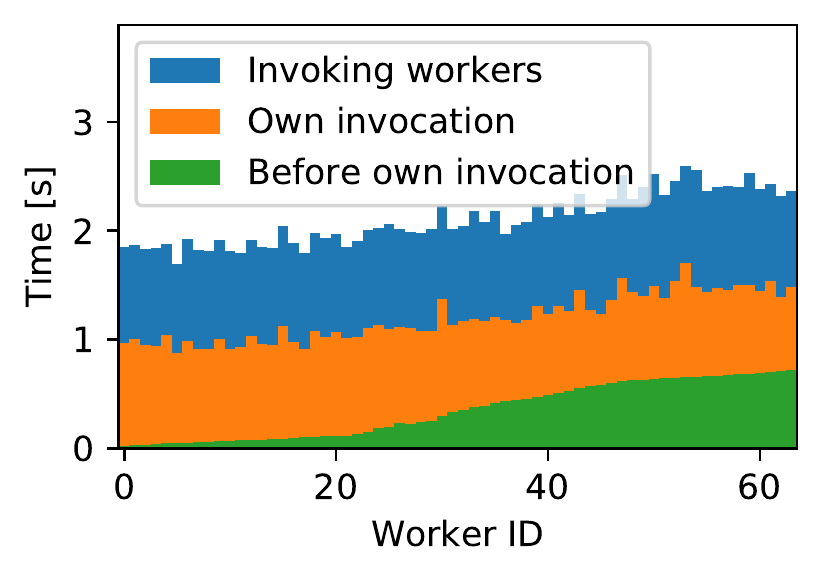}
  \caption{Example run of two-level invocation process.}
  \label{fig:example-startup}
\end{figure}

Figure~\ref{fig:example-startup} shows the timings
of an example run using this approach to start 4096 serverless workers
based on a freshly created function (i.e., performing a cold start).
It shows a timeline with three phases of every first-generation worker
in the order they are invoked by the driver:
(1) the time the driver took before it initiated their invocation
(namely, to launch all previous workers),
(2) the time their invocation took, i.e.,
the time between their invocation was initiated and they were actually running,
and (3) the time they took to do the second-generation invocations.
As the plot shows, the invocation of the last worker
was initiated after about \SI{2.5}{\second},
which is tremendously faster than the \SIrange{13}{18}{\second}
that the driver would be expected to take for doing the invocations alone
based on the invocation rates from Table~\ref{tbl:invocation-characteristics}.

Note that the limit on the invocation rate of AWS Lambda is not relevant:
it is currently ten times
the limit on the number of concurrent invocations (i.e., workers) per second.
Each query only needs \emph{one} invocation per worker
and the single user of our function is expected to run queries
at a rate orders of magnitudes lower than ten per second.
The limit on concurrent invocations, however, \emph{is} relevant
and we discuss some more details in Section~\ref{sec:experiments}.

\subsection{S3-based Scan Operator}
\label{subsec:s3_storage_analysis}

\subsubsection{Network Characteristics}

\begin{figure}
    \centering
    \subcaptionbox{Large files (\SI{1}{\giga\byte}).
                   \label{fig:scan-large-files}}{
        \includegraphics[height=.41\linewidth]{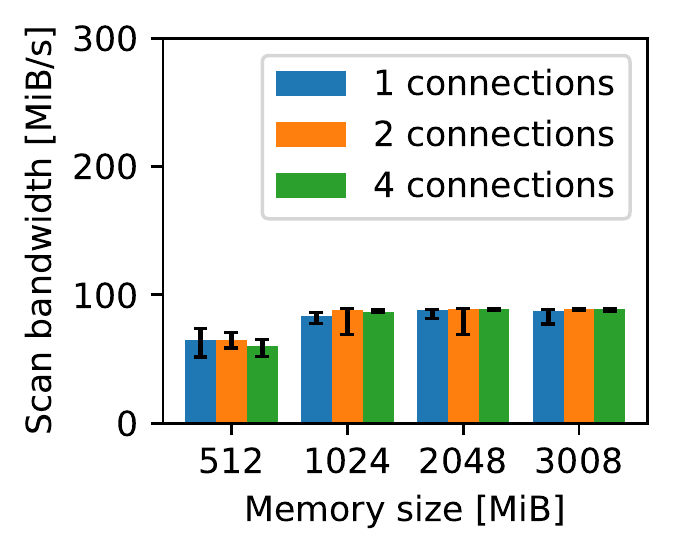}}
    \subcaptionbox{Small files (\SI{100}{\mega\byte}).
                   \label{fig:scan-small-files}}{
        \includegraphics[height=.41\linewidth]{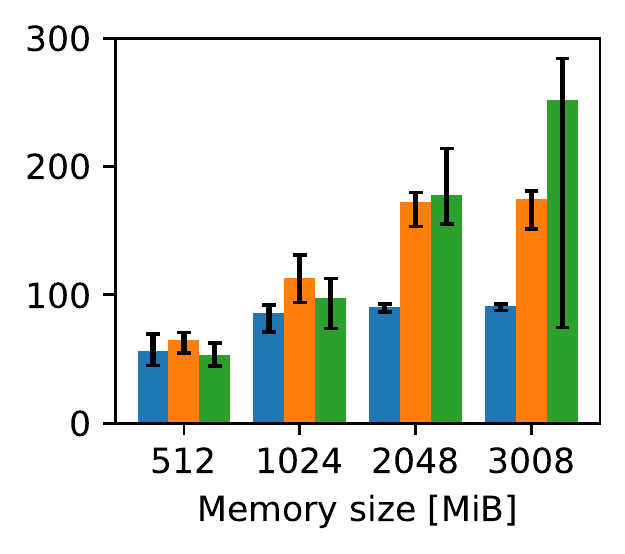}}
    \caption{Network (ingress) bandwidth of serverless workers.}
    \label{fig:scan-file-sizes}
\end{figure}

We first study the characteristics in terms of performance and cost
of accessing S3 from the serverless workers
in order to derive design principles for implementing scan operators.
Figure~\ref{fig:scan-file-sizes} presents microbenchmarks
for downloading large and small files from S3
into serverless workers using different configurations.
We run each configuration three times in direct succession
in nine different data centers using ten workers in each run.
We compute the median, minimum, and maximum bandwidth
of all workers in each data center
and plot the medians of the three values as the colored bars, the lower error,
and the upper error, respectively.
The plots thus show the distribution in a ``median'' data center.

For large files (Figure~\ref{fig:scan-large-files}),
there is a very stable limit of about \SI{90}{\mebi\byte/\second} per worker.%
\footnote{This is about \factor{2} higher
          than the numbers reported by \textcite{pywren} published in 2017.
          It is also qualitatively different
          from the results of \textcite{Wang2018},
          who reported a stronger correlation between
          network bandwidth and worker memory size published 2018.
          We assume that Amazon has increased the network bandwidth since then.}
Workers of virtually any size have fast enough network to achieve this limit;
only workers with less than \SI{1}{\giga\byte} of main memory
see a slightly lower ingress bandwidth.
Furthermore, using more network connections
does not significantly change the overall bandwidth.

For small files (Figure~\ref{fig:scan-small-files}),
the picture looks different.
Workers with large amounts of memory observe a much higher network bandwidth,
occasionally reaching almost \SI{300}{\mebi\byte/\second}.
However, this is only the case
if they use several network connections at the same time.
We do not have access to information about Amazon's network infrastructure,
but we assume that it uses a credit-based traffic shaping mechanism
to limit the network bandwidth of each function instance
to the \SI{90}{\mebi\byte/\second} observed above.
Such a mechanism would allow bursts to exceed the target limit
for a short amount of time and thus explain our results.
In experiments not shown, we observe that the time span
during which the burst may exceed the target is a small number of seconds.
In order to maximize performance for short-running scans,
we thus need to use multiple concurrent connections.

\begin{figure}
  \centering
  \includegraphics[width=.8\linewidth]{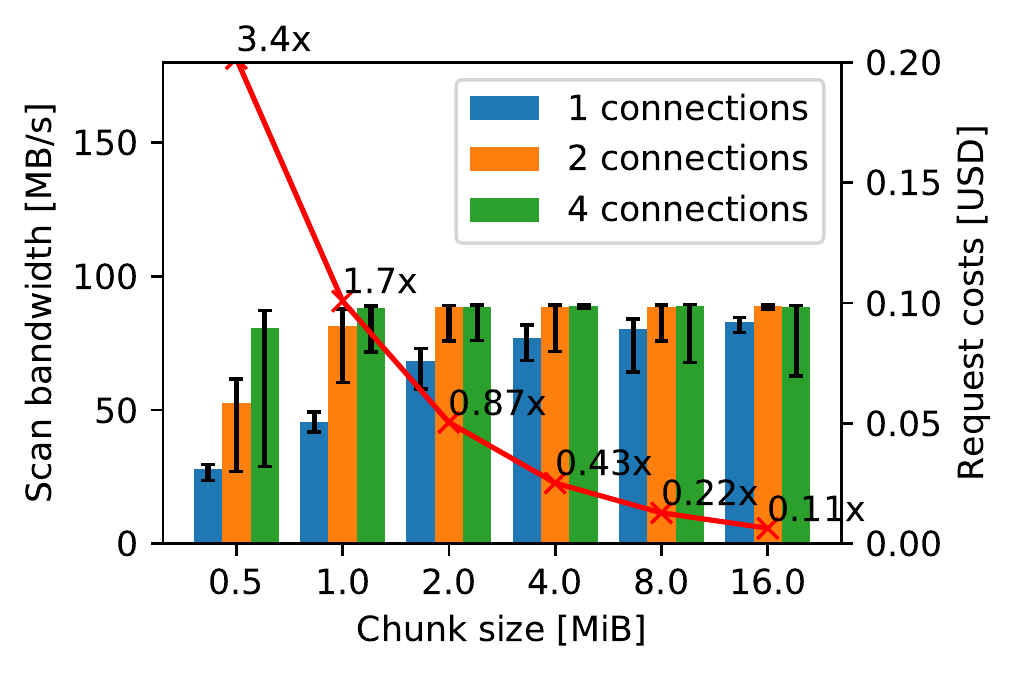}
  \caption{Impact of the chunk size on scan characteristics.}
  \label{fig:scan-chunksize}
\end{figure}

We also study the impact of the size of each individual request to S3
on the bandwidth and cost of a scan.
To that aim, we download a file of \SI{1}{\giga\byte}
with requests of different sizes using a variable number of connections.
Figure~\ref{fig:scan-chunksize} shows the result
for the largest available serverless workers
(i.e., with \SI{3008}{\mebi\byte} of main memory).
While a single connection requires a chunk size of \SI{16}{\mega\byte}
to get reasonably close to the maximum throughput from the previous experiment,
we achieve that throughput even with a chunk size of \SI{1}{\mega\byte}
using four concurrent connections.
This is the classical technique of hiding the latency of one or more requests
with the processing of another.
The size of each request also has a direct impact
on the overall costs of a scan:
it is inversely proportional to the \emph{number} of requests,
each of which has a fixed cost.
One million read requests currently cost%
\footnote{In the ``us-east-1'' region,
          see \url{https://aws.amazon.com/s3/pricing/\#Request\_pricing}.}
\SI{0.4}[\$]{}.
The line in Figure~\ref{fig:scan-chunksize} shows
the costs of running the experiment one thousand times.
It is annotated with the factor
by which the requests are more expensive than running the serverless workers.
For example, in a scan with a chunk size of \SI{1}{\mebi\byte},
the requests are \SI{1.7}{\times} more expensive
than the workers cost for the same scan.
With even smaller chunk sizes,
the requests can easily dominate the overall cost.
In order to support small reads from S3,
we thus need to support several in-flight requests,
but also avoid small reads wherever possible.

\subsubsection{Operator Design}

We use above insights to design a scan operator
that uses the network and CPU resources efficiently.
We describe the design of a scan operator for Parquet files as an example,
but expect the design of other operators to be conceptually similar.
Parquet files are not only well-suited
because they are wide-spread and optimized for slow storage,
but they are also configurable in several ways
such that they exhibit many characteristics that other formats might have.

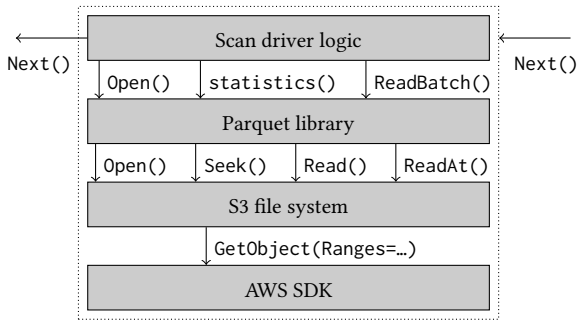
\begin{figure}
  \centering
  \input{figures/scan-operator}
  \caption{Components of the Parquet scan operator.}
  \label{fig:scan-operator}
\end{figure}

Figure~\ref{fig:scan-operator} shows the main components of the operator.
To the outside, it implements the open/next/close operator interface,
through which it reads one or more file paths from its upstream operator
and returns their content to its downstream operator
as a sequence of table chunks in columnar format.
In a typical plan, these chunks are consumed by a JiT-compiled pipeline,
whose first operator is a scan operator for in-memory table chunks,
which extracts individual records.
Internally, the Parquet scan operator
uses the official C++ library for Parquet files%
\footnote{The C++ library for Parquet is part of Apache Arrrow,
          see \url{https://github.com/apache/arrow}.}
to handle the deserialization of data and metadata.
We have implemented the user-level filesystem interface of that library
with a backend for S3, which, in turn, uses Amazon's AWS SDK for C++
to make requests to the S3 REST endpoint over the network.

The Parquet format has been designed to enable
pushing down selections and projections.
To that aim, the data is stored
in consecutive groups of rows called \emph{row groups},
each of which stores its records
as consecutive columns called \emph{column chunks}.
Each column chunk may use a light-weight and a heavy-weight compression scheme,
such as run-length encoding and GZIP, respectively.
Furthermore, the footer of the file contains (optional) min/max statistics
as well as absolute offsets for each column chunk.
The library loads this metadata with a single file read,
exposes the statistics to the scan operator
such that it prunes out row groups based on its predicates,
and loads the column chunks of the projected attributes
when the scan operator accesses the remaining row groups
using read operation per column chunk.
Each of these read operations on the file system
is translated to one request to S3,
which downloads the desired bytes of the file
(using HTTP's \texttt{Ranges} header).
The file system offers a random-access interface
(through \texttt{ReadAt},
as opposed to a stream-like interface through \texttt{Seek} and \texttt{Read})
which supports multiple concurrent reads.

We identify four levels where concurrent connections could be used
to maximize bandwidth utilization on small files and small chunks,
thus addressing the insights from Figures~\ref{fig:scan-small-files}
and \ref{fig:scan-chunksize}, respectively:
(1) making several requests for each read operation in the filesystem,
(2) downloading different column chunks of the same row group,
(3) downloading multiple row groups at the same time,
and (4) downloading data or metadata from different files at the same time.
We always exploit level (4) by consuming the list of paths eagerly
and downloading the metadata for all files that should be scanned
in a dedicated thread in order to hide the latency of these small requests.
Next, we exploit level (3) by downloading
the data of up to two row groups asynchronously in two dedicated threads,
except if the worker has too little main memory.
This also overlaps the download(s) of one row group
with the decompression and subsequent processing of the previous one.
For small files and files with a single row group, we exploit level (2)
by downloading different column chunks using multiple threads.
We only fall back to level (1) if none of the other levels could be exploited
as this would increase the number of requests and thus the costs of a scan.
We expect that a similar prioritization to be applicable to other formats as well.

Finally, we exploit the (small amount of) multi-core parallelism
in the workers with more than \SI{1792}{\mebi\byte} of memory
(see Figure~\ref{fig:compute-throughput})
by optionally parallelizing the decompression of column chunks.
This is only beneficial
if decompressing a column chunk is slower than downloading it,
which is only the case for the most heavy-weight compression schemes,
and if the remaining query has too little compute
to utilize the resources fully.

\subsection{Exchange Operator}
\label{sec:exchange-operator}

As one major building block for data processing,
we design a family of exchange operators for serverless workers.
Since serverless workers cannot accept incoming connections,
they can only communicate through external storage.
In order to support exchanging large amounts of data,
we use S3 for this purpose.

\subsubsection{Basic Ideas and Challenges}

\begin{algorithm}
  \begin{algorithmic}[1]
    \Function{BasicExchange}{$p$\textbf{: Int}, $\mathcal{P}$\textbf{: Int}$[1..P]$, $R$\textbf{: Record}$[1..N]$,\par%
                             \hspace{2em}\textsc{FormatFileName}\textbf{: Int \texttimes{} Int $\rightarrow$ String}}%
      \State partitions $\gets$ \Call{DramPartitioning}{$R$, $\mathcal{P}$}
      \label{algl:basic-exchange-partitioning}
      \For{$\langle \varname{receiver}, \text{data} \rangle$ \textbf{in} partitions}
        \State \Call{WriteFile}{\textsc{FormatFileName}($\varname{receiver}$, $p$), data}
      \EndFor
      \For{$\varname{source}$ \textbf{in} $\mathcal{P}$}
        \State data $\gets$ data $\cup$ \Call{ReadFile}{\textsc{FormatFileName}($p$, $\varname{source}$)}
      \EndFor
      \State \textbf{return} data
    \EndFunction
  \end{algorithmic}
  \caption{Basic S3-based exchange operator.}
  \label{alg:basic-exchange}
\end{algorithm}

Algorithm~\ref{alg:basic-exchange} shows how the basic exchange algorithm works,
which other authors have used as well%
~\cite{pywren,Klimovic2018,Pu2019}:
Each worker $p$ of the $P$ workers holds its share $R$ of the input relation
and uses an in-memory partitioning routine
to split its input into $P$ partitions,
for example based on the hash value of their key.
It then writes the data of each partition into a file
whose name reflects its own ID
as well as the ID of the ``receiver'' of the file.
Finally, it reads all files where its own ID has been used as receiver
sent by any of the other ``source'' workers.
Since the sender may be slower than the receiver,
the receiver must repeat reading a file until that file exists.

The problem with this algorithm is
that the total number of files is quadratic in $P$:
each of the $P$ workers reads from and writes to $P$ files.
This may cause throttling by the cloud provider
due to a rate limit on requests.
For 1k workers, one execution of \textsc{BasicExchange} needs 2M requests
while, as of July 2018, the rate limit on AWS is 3.5k and 5.5k per second
for writes and reads, respectively,%
\footnote{See \url{https://aws.amazon.com/about-aws/whats-new/2018/07/amazon-s3-announces-increased-request-rate-performance/}}
and was as low as 300 and 800 read and write requests per second before that.%
\footnote{See \url{https://forums.aws.amazon.com/message.jspa?messageID=573975\#573975}.}
This effect has been pointed out by previous work~\cite{Klimovic2018,pywren},
which solved the problem
by running their own storage service on rented VM instances.

However, it is possible to by-pass the rate limits with a simple trick:
By implementing \textsc{FormatFileName} such that it encodes
the sender and/or receiver ID or parts thereof in the \emph{bucket name},
we increase the overall rate limit by the number of buckets we use.
For example, if we use:
\begin{equation*}
\begin{split}
&\textsc{FormatFileName} := \langle s, r \rangle \mapsto \\
            & \qquad \text{``\texttt{s3://bucket-\linebreak\{$r \% 10$\}/sender-\{$s$\}/receiver-\{$r$\}}''}
\end{split}
\end{equation*}
We reduce the rate request to $P/10$ requests per second per bucket,
which is below the historic limits for up to 3k workers.
This requires to create $10$ buckets,
but this can be done at installation time and does not induce costs.

While this scheme solves the performance problem,
it incurs prohibitive costs,
which also grow quadratically in the number of workers.
As of writing, 1M read and write requests cost
\SI{5}[\$]{} and \SI{.4}[\$]{}, respectively.%
\footnote{In the ``us-east-1'' region,
          see \url{https://aws.amazon.com/s3/pricing/\#Request\_pricing}.}
The left-most bars (labeled \textsc{1l}) in Figure~\ref{fig:exchange-cost} show
how the cost of \textsc{BasicExchange} evolves with the number of workers.
With 256 workers, the costs for the requests to S3
are already higher than the costs
for running the workers in most typical configurations,
which are indicated by the horizontal range.
With 4k workers, running the algorithm on \SI{4}{\tebi\byte}
costs about \SI{100}[\$]{} for the requests to S3
and \SI{3.3}[\$]{} for running the workers.

In the remainder of this section,
we present two orthogonal optimizations
that reduce the number of requests.

\subsubsection{Multi-Level Exchange}

The first idea is to do the exchange in multiple levels,
where each level only involves a small subset of the workers.

For two levels, we project the partition and worker IDs onto a grid
and first do a horizontal exchange and then a vertical exchange.
To that aim, we define the projection function
$H_s := \langle H^1_s, H^2_s \rangle := x \mapsto \langle x\,\%\,s, x \sslash s\rangle$,
which projects a number $x \in \{1..P\}$ onto two coordinates,
where $s$ is the desired number of distinct elements in the first dimension
and $ \%$ and $\sslash$ are modulo and integer division, respectively.
Note that this approach works also for non-quadratic numbers of workers $P$.
As a building block, we use \textsc{BasicGroupExchange},
which is the \textsc{BasicExchange} as defined before
extended by a parameter for a projection function $H_i$,
which it applies to the partition IDs
while running the in-memory partitioning routine
(Line~\ref{algl:basic-exchange-partitioning}
in Algorithm~\ref{alg:basic-exchange}).

\begin{algorithm}
  \begin{algorithmic}[1]
    \Function{TwoLevelExchange}{$p$\textbf{: int}, $P$\textbf{: int}, $R$\textbf{: Record} $[1..N]$}%
      \State $\langle p_1, p_2 \rangle \gets  H_s(p)$
      \State $\mathcal{P}_i \gets \{q|q\in\{1..P\}:q_i = p_i\}$ \textbf{for} $i=1,2$
      \State $f_i \gets \langle s, t \rangle \mapsto$ ``\texttt{s3://b\{$p_i$\}/snd\{$s$\}/rcv\{$r$\}}'' \textbf{for} $i=1,2$
      \State tmp $\gets$ \Call{BasicGroupExchange}{$p$, $\mathcal{P}_1$, $f_1$, $R$, $H^2_s$}
      \State \textbf{return}$\,$\Call{BasicGroupExchange}{$p$, $\mathcal{P}_2$, $f_2$, tmp, $H^1_s$}
    \EndFunction
  \end{algorithmic}
  \caption{Two-level S3-based exchange operator.}
  \label{alg:twolevel-exchange}
\end{algorithm}

Algorithm~\ref{alg:twolevel-exchange} shows
how the two-level exchange works.
We first compute the two-dimensional worker ID from $p$.
We then define the set of coworkers $\mathcal{P}_1$
that have the same value in the first coordinate
and run \textsc{BasicGroupExchange} on the input
to exchange data with this subset of workers.
To do so, we parameterize the routine the following way:
First, we use $f_1$ as \textsc{FormatFileName},
which prefixes all file names with a distinct bucket of this group,
``\texttt{s3://b\{$p_1$\}}''.
Second, we have it apply the projection function $H^2_s$ to the partition ID,
which means that it considers the second coordinate of the IDs
for this round of the exchange.
When this function returns,
the second coordinate of the partition ID of any record
coincides with that of the worker ID where it resides.
Finally, we reverse the roles of the first and second coordinate
and run \textsc{BasicGroupExchange} again,
now among the group of workers induced by the other coordinate, $\mathcal{P}_2$.
After this step, the first coordinate of the partition ID of any record
also coincides with that of the worker ID where it resides,
so the exchange is complete.

The two-level approach reduces the number of requests,
as the number of each phase grows only quadratically
with the \emph{group size} instead of the number of workers.
More precisely, each worker does $P/s$ read and write requests
in the first level and, by definition, $s$ in the second.
Thus, together the $P$ workers do $P^2/s$ and $Ps$ requests
in the first and second level, respectively.
It is easy to see that $s = \sqrt{P}$ minimizes the sum of the two terms,
so we use this value for the rest of the paper.
In total, the algorithm does hence $2P\sqrt{P}$ read and write requests each.
At the same time, it reads and writes the input two times instead of just one,
which increases run time and hence the cost of running the workers.
We study this trade-off in more detail below.
Finally, the number of requests \emph{per bucket},
which is the metric that is subject to the rate limits,
is $P\sqrt{P}/B$ per round
($P$ workers spreading $\sqrt{P}$ requests each over $B$ buckets).
Theoretically, one round of exchange
with \SI{10}{\kilo\nothing} workers and \num{300} buckets
should thus take at most \SI{3}{\second} under the current limits.

The same idea can be applied to three or more levels
to reduce the number of requests even further.
For $k$ levels, the partition and worker IDs
are projected onto a $k$-dimensional grid with side length $\sqrt[k]{P}$
and \textsc{BasicGroupExchange} is used $k$ times, once for each dimension
(each of which reads and writes the input once).
Table~\ref{tbl:exchange-cost-model} summarizes
the characteristics of the different algorithms.

\begin{table}
  \caption{Cost models of S3-based exchange algorithms.}
  \label{tbl:exchange-cost-model}
  \centering
  \begin{tabular}{@{}lcccc@{}}
    \toprule
    \textbf{Algorithm} & \textbf{\#reads} & \textbf{\#writes} & \textbf{\#lists} & \textbf{\#scans} \\
    \midrule
    1l                 & $P^2$            & $P^2$             & $\Oh(P)$         & 1                \\
    1l-wc              & $P^2$            & $P$               & $\Oh(P)$         & 1                \\
    \midrule
    2l                 & $2P\sqrt{P}$     & $2P\sqrt{P}$      & $\Oh(P)$         & 2                \\
    2l-wc              & $2P\sqrt{P}$     & $2P$              & $\Oh(P)$         & 2                \\
    \midrule
    3l                 & $3P\sqrt[3]{P}$  & $3P\sqrt[3]{P}$   & $\Oh(P)$         & 3                \\
    3l-wc              & $3P\sqrt[3]{P}$  & $3P$              & $\Oh(P)$         & 3                \\
    \bottomrule
  \end{tabular}
\end{table}

\subsubsection{Write Combining}

The second idea consists in
writing all partitions produced by one worker into a single file.
We call this optimization ``write combining''.
Instead of reading one entire file per sender,
the receivers now need to read \emph{part of one file} per sender.
We thus define \textsc{FormatFileName} such that
it ignores the parameter value for the receiver.

To communicate the offsets of each part to the receivers,
we consider two variants:
we either write the offsets into a separate file,
which doubles the amount of read requests,
or we encode the offsets into the file name.
In the second variant,
we change \textsc{FormatFileName} to accept an additional parameter \emph{offsets}
which it appends to the end of each file name.
The receivers now do not know the names of the files they should receive.
However, they can find out the filenames using a \emph{list} request
(which they may need to repeat a few times
until they see the files produced by all senders).
This way, the receivers can read the offsets of all senders
using one or very few requests.
Currently, AWS charges list requests for the price of write requests,
so, in addition to the potential performance gain,
this variant is cheaper for more than about 12 workers.
On the other hand, file names are limited to \SI{1}{\kibi\byte},
so this only works until at most a few hundred workers
depending on the offset granularity, maximum offset, and encoding scheme,
but this is enough for the multi-level variants.%
\footnote{For example, in the two-level variant,
          10k workers are split into groups of 100.}

\subsubsection{Serverless Exchange Operator Analysis}
\label{subsubsec:exch_op_analysis}

\begin{figure}[t]
  \centering
  \includegraphics[width=\linewidth]{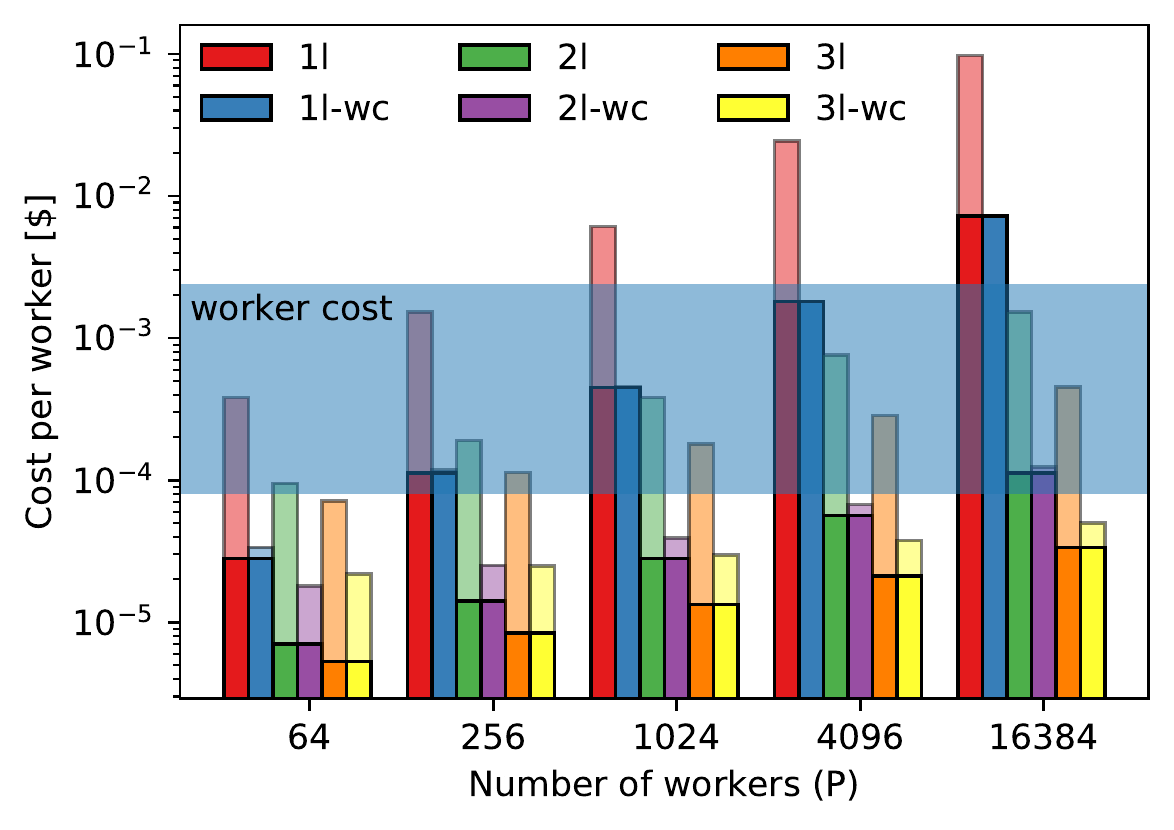}
  \caption{Cost of S3-based exchange algorithms on AWS.}
  \label{fig:exchange-cost}
\end{figure}

In Figure~\ref{fig:exchange-cost},
we compare the costs of the different algorithms.
Here, we show $i$ exchange levels with and without write combining ($wc$).
To compute the costs for the requests,
we use the cost models from Table~\ref{tbl:exchange-cost-model}
at the rates quoted above
(\SI{5}[\$]{} and \SI{.4}[\$]{}
for 1M read and write requests cost, respectively).
The lower bars in full color represent the read cost,
the upper bars in lighter color represent the write cost
of the respective algorithms.
\textsc{BasicExchange} is labelled \textsc{1l},
while the two- and three-level variants
are labelled \textsc{2l}, and \textsc{3l};
variants using write combining are suffixed with \textsc{-wc}.
The horizontal range shows the costs for running the workers.
For the purpose of this plot,
we assume that they achieve \SI{85}{\mebi\byte/\second},
do not experience waiting time,
and each second costs \SI{3.3e-5}[\$]{}
(which is the current price on AWS for workers with \SI{2}{\gibi\byte} RAM).
The lower edge of the range represents the running costs
of the workers doing one scan on an input of \SI{100}{\mebi\byte}
while the upper edge represents the costs for three scans of \SI{1}{\gibi\byte}.
This range helps to put the costs of the requests into perspective.

As observed before, the plot shows that
the costs of \textsc{BasicExchange} do not scale with the number of workers.
While using write combining reduces the write costs to a negligible amount,
the read cost, which still grow quadratically,
can still be dominant in many cases.
Using two levels has always lower request costs than using just one,
and, when used with write combining,
reduces the costs of all requests of an exchange
below the worker costs in almost all configurations.
Using three levels and write combining brings them to a negligible level
in all configurations considered here.

Overall, the two optimizations give us effective knobs
to reduce the costs due to requests to storage.

%% file: figures/scan-operator.tex
\newlength\mctextheight
\setlength\mctextheight{\heightof{$\vec{C}\vec{S}$}}
\begin{tikzpicture}[
      scale=.9,transform shape,
      every node/.style={
        text height=\mctextheight,text depth=depth("y$a_0$"),
        node distance=3.5em,font=\small},
      layer/.style={
        draw,fill=white!80!black,minimum width=.7\linewidth,},
    ]
  \node [layer]
    (driver) {Scan driver logic};
  \node [layer,below=of driver.south east,anchor=south east]
    (parquet) {Parquet library};
  \node [layer,below=of parquet.south east,anchor=south east]
    (s3fs)   {S3 file system};
  \node [layer,below=of s3fs.south east,anchor=south east]
    (awssdk) {AWS SDK};
  \node [draw,densely dotted,fit=(driver)(parquet)(s3fs)(awssdk)]
    (operator) {};

  \draw [->] ($(driver.south west)+(.5em,0)$) --
             ($(parquet.north west)+(.5em,0)$)
    node [draw=none,midway,anchor=west] (parquet-open) {\texttt{Open()}};
  \draw [->] ($(parquet-open.east |- driver.south)+(.85em,0)$) --
             ($(parquet-open.east |- parquet.north)+(.85em,0)$)
    node [draw=none,midway,anchor=west] (parquet-stats) {\texttt{statistics()}};
  \draw [->] ($(parquet-stats.east |- driver.south)+(.85em,0)$) --
             ($(parquet-stats.east |- parquet.north)+(.85em,0)$)
    node [draw=none,midway,anchor=west] (parquet-read) {\texttt{ReadBatch()}};

  \draw [->] ($(parquet.south west)+(.35em,0)$) --
             ($(s3fs.north west)+(.35em,0)$)
    node [draw=none,midway,anchor=west] (s3fs-open) {\texttt{Open()}};
  \draw [->] ($(s3fs-open.east |- parquet.south)+(.8em,0)$) --
             ($(s3fs-open.east |- s3fs.north)+(.8em,0)$)
    node [draw=none,midway,anchor=west] (s3fs-seek) {\texttt{Seek()}};
  \draw [->] ($(s3fs-seek.east |- parquet.south)+(.8em,0)$) --
             ($(s3fs-seek.east |- s3fs.north)+(.8em,0)$)
    node [draw=none,midway,anchor=west] (s3fs-read) {\texttt{Read()}};
  \draw [->] ($(s3fs-read.east |- parquet.south)+(.8em,0)$) --
             ($(s3fs-read.east |- s3fs.north)+(.8em,0)$)
    node [draw=none,midway,anchor=west] (s3fs-readat) {\texttt{ReadAt()}};

  \draw [->] ($(s3fs.south west)+(5em,0)$) --
             ($(awssdk.north west)+(5em,0)$)
    node [draw=none,midway,anchor=west] (awssdk-open)
    {\texttt{GetObject(Ranges=\hbox to.5em{\hss.\hss\hss.\hss\hss.\hss})}};

  \draw [->] (driver.west) --
             ($(driver.west)-(3em,0)$)
    node [draw=none,midway,anchor=north,xshift=-.5em] {\texttt{Next()}};
  \draw [<-] (driver.center -| operator.east) --
             ($(driver.center -| operator.east)+(3em,0)$)
    node [draw=none,midway,anchor=north,xshift=.5em] {\texttt{Next()}};
\end{tikzpicture}

%% file: experiments.tex
\section{Evaluation}
\label{sec:experiments}

\subsection{Dataset and Methodology}

In most experiments, we use the \texttt{LINEITEM} relation from TPC-H,
which we generate at scale factor (SF) 1000.
Since our prototype does not support strings yet,
we modify \texttt{dbgen} to generate numbers instead of strings.
Furthermore, we sort the \texttt{LINEITEM} relation by \textit{l\_shipdate}
in order to show the effect of selection push downs on that attribute.
In uncompressed CSV, the size of the relation is \SI{705}{\gibi\byte};
in Parquet using standard encoding and GZIP compression,
the size is \SI{151}{\gibi\byte}.
Following the best practices of big data processing
and the systems we compare to below,
we store the Parquet data in files of about \SI{500}{\mega\byte}.
In order to generate data at higher scale factors
within a reasonable amount of time,%
\footnote{\texttt{dbgen} took one week on our ten-machine cluster
          to generate SF 1000.}
we replicate the files of SF~1000 accordingly,
which should preserve query properties.

Unless otherwise mentioned,
we measure the end-to-end query latency, which accounts for the
serverless workers invocation time, the useful work carried out, and fetching
the results from the result queue in Amazon SQS.
We report the median of three runs in the same data center,
as we observed little variation across data centers
in the experiments shown in Figures~\ref{fig:scan-file-sizes}
and \ref{fig:scan-chunksize},
as well as other isolated experiments not shown
(with the exception of invocation into the cloud,
as shown in Table~\ref{tbl:invocation-characteristics}).
Since the default limit
of concurrent function executions is \SI{1}{\kilo\nothing},
we had to increase this limit through a support request,
which is possible without further cost and was handled within less than a day.

\subsection{Effect of Worker Configurations}
\label{subsec:eval_swarm_size}

In this experiment, we explore the parameter space of worker configurations to
gain a deeper understanding of their impact.
Specifically, we vary the amount of main memory of each worker, $M$,
which influences the amount of CPU cycles the function can use,
as well as the number of files, $F$, that each worker processes.
The latter parameter indirectly defines the number of workers:
the tables is stored in \si{320} files, so we use $W=320/F$ workers.
We use TPC-H Query~1 (Q1),
which selects \SI{98}{\percent} of \texttt{LINEITEM}
and aggregates them to a very small amount of groups,
in order to eliminate effects of more complex plans.
We create a fresh function for each configuration and each repetition
and run the query twice,
the first one as a \emph{cold} run, the second as a \emph{hot} run.

\begin{figure*}
    \centering
    \begin{subfigure}[t]{.2265\linewidth}
        \centering
        \includegraphics[width=\linewidth]{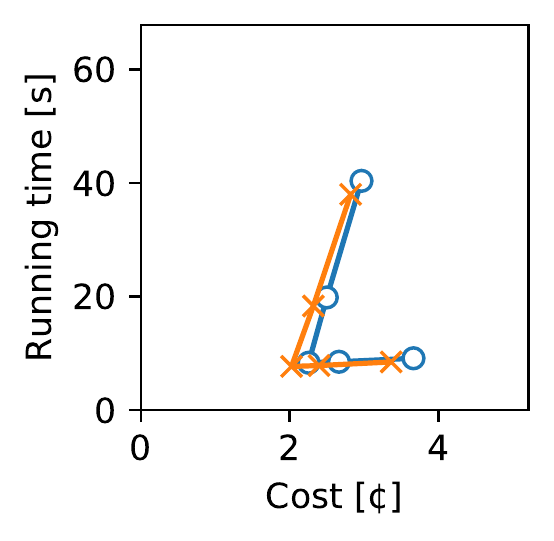}
        \caption{$F = 1$, varying $M$.}
        \label{fig:eval:tpch-sf1k-q1-mem}
    \end{subfigure}
    ~
    \begin{subfigure}[t]{.207\linewidth}
        \centering
        \includegraphics[width=\linewidth]{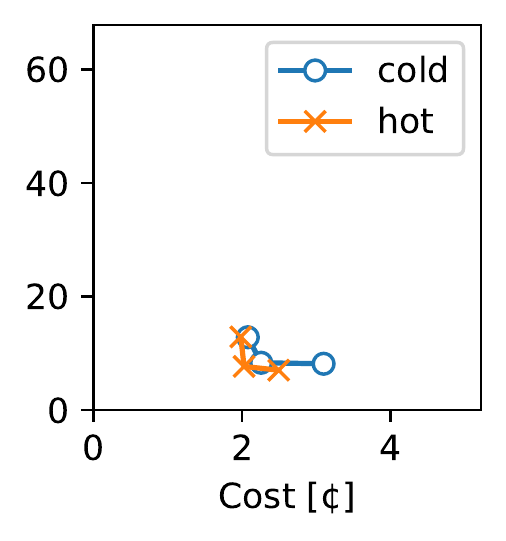}
        \caption{$M\hspace{-.2em}=\hspace{-.2em}\SI{1792}{\mebi\byte}$, varying $F$.}
        \label{fig:eval:tpch-sf1k-q1-files}
    \end{subfigure}
    ~
    \begin{subfigure}[t]{.207\linewidth}
        \centering
        \includegraphics[width=\linewidth]{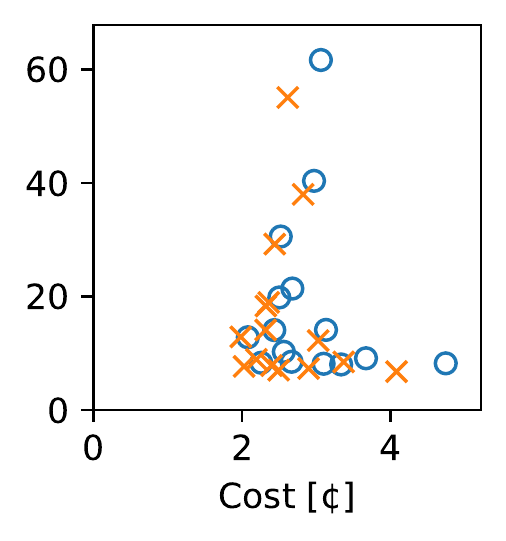}
        \caption{Varying $M$ and $F$.}
        \label{fig:eval:tpch-sf1k-q1-all}
    \end{subfigure}
    \caption{TPC-H Query 1 with varying memory ($M$)
             and number of files ($F$) per worker.}
    \label{fig:eval:tpch-q1}
\end{figure*}

Figure~\ref{fig:eval:tpch-q1} shows the result.
First, we fix the number of files per worker to $F=1$ (i.e., $W=320$)
vary the memory size allocated per worker
(\si{512}, \si{1024}, \si{1796}, \si{2048}, and \SI{3008}{\mebi\byte}).
As Figure~\ref{fig:eval:tpch-sf1k-q1-mem} shows,
by increasing the worker size from \si{512} to \SI{1796}{\mebi\byte},
execution gets significantly faster.
This is because scanning GZIP-compressed data is CPU-bound
and more memory means more CPU cycles
as described in Section~\ref{subsec:intraworker_parallelism}.
Interestingly, it also gets marginally cheaper.
We attribute this to the overhead of multi-threading
in a configuration where that does not yield any gains
and thus only reduces efficiency.
As we increase the worker size further,
the price increases (due to the linear relationship in the price model),
however, without improving speed.
Similar to related work, we observe a small penalty
on the end-to-end latency of cold runs of about \SI{20}{\percent}.
This is not only due to a slower invocation time,
but also somewhat slower execution
(possibly due to loading of code from the dependency layer),
which affects the price.
Despite of that, both hot and cold execution
return in less than \SI{10}{\second}
and thus within a timeframe that is suitable interactive analytics.

Figure~\ref{fig:eval:tpch-sf1k-q1-files} shows the results
for varying the number of files per worker $F=\{4,2,1\}$,
and with it the number of workers $W=\{80,160,320\}$,
while fixing the worker size to $M=\SI{1796}{\mebi\byte}$.
This is essentially the same experiment
as the simulation from the Introduction
shown in Figure~\ref{fig:intro-job-scoped-resources}:
using more workers speeds up execution,
but at diminishing gains and thus increased costs.
Finally, Figure~\ref{fig:eval:tpch-sf1k-q1-all}
shows all different combinations of $M$ and $F$.
Which of the configurations (on the pareto-optimal front)
a user might want to pick depends on her preference for price or speed
and is out-of-scope of this paper.
In the remainder of the paper, we either manually pick a good trade-off
or show a range of configurations.

\subsection{Effect of Push-downs}
\label{subsec:effect_selectivity}

In order to study the effect of pushing down
selections and projects into the scan operator,
we run the two most scan-bound queries from TPC-H, Query~1 and 6.
While Query~1 selects \SI{98}{\percent} of the relation
and uses seven attributes,
Query~6 selects only \SI{2}{\percent} of it, but uses four attributes.
In order to eliminate unrelated effects such as invocation time,
we only measure processing time in this experiment,
i.e., the time each worker takes to executes its plan fragment.

\begin{figure}
  \centering
  \includegraphics[width=.7\linewidth]{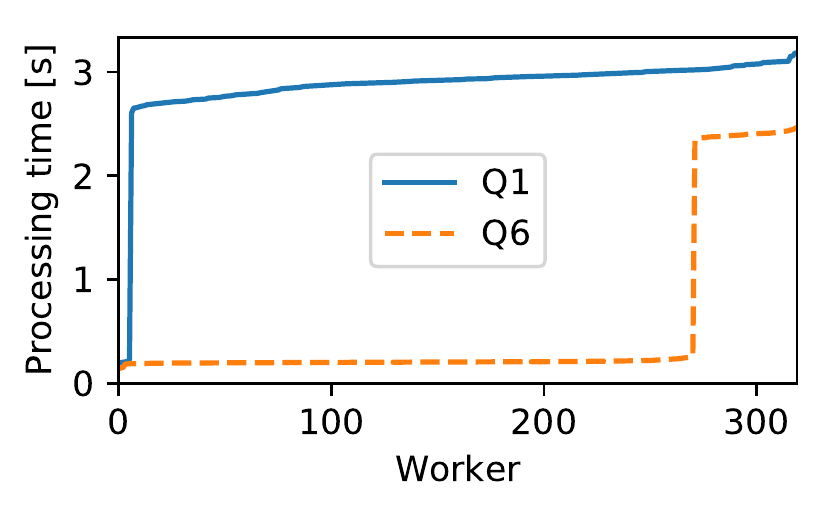}
  \caption{Distribution of processing time.}
  \label{fig:eval:tpch-sf1k-runners-duration}
\end{figure}

Figure~\ref{fig:eval:tpch-sf1k-runners-duration}
shows the processing time of all workers
ordered by increasing processing time
using $F=1$ and $M=\SI{1792}{\mebi\byte}$.
In both queries, there are two categories of workers:
those where the processing time is \SIrange{100}{200}{\milli\second}
and those where it is \SIrange{2}{3}{\second}.
The workers of the former category loads the metadata of their file
(inducing one round-trip to S3),
prune out all row groups due to the min/max indices on \texttt{l\_shipdate},
and immediately return an empty result.
For Query~1, this happens to about \SI{2}{\percent} of the workers;
for Query~6, to about \SI{80}{\percent},
which corresponds to the respective selectivity
of the filter on \texttt{l\_shipdate}.
If the min/max indices were stored in a central place
and available before starting the workers,
these workers would not even be started,
but such optimizations are out of the scope of this paper.
The other workers cannot prune out anything,
so they load the projected columns from S3 and decompress and scan them.
For them, the data volume of the projected columns determine the execution time,
which is slightly higher in Query~1 than in Query~6.

\subsection{Comparison with \qaas Systems}
\label{subsec:lambada_qaas_systems}

\subsubsection{Experiment Setup}

We compare Lambada with two \Qaasl systems,
\bigquery~\cite{big_query} and \athena~\cite{athena}.
This type of system has a similar operational simplicity as Lambada,
namely the ability to query datasets from cloud storage
without starting or maintaining infrastructure,
and is thus the most similar alternative
for interactive analytics on cold data.

In practice, only \athena supports
\emph{in-situ} processing of large-scale datasets.
\bigquery can currently only process individual files without prior loading
(which is subject to further restrictions).
Large-scale datasets need to be loaded with an ETL process,
during which they are converted into a proprietary data format
and possibly indexed.
In this format, our \texttt{LINEITEM} table takes \SI{823}{\gibi\byte},
which is slightly larger than the uncompressed CSV
and over \SI{5}{\times} larger than our Parquet files.
The promise of loading into this format is to allow for faster querying.
We still include \bigquery in our study
as the system otherwise fits well
and the cloud provider could lift this restriction in the future.
For \athena, we use the same files as for \lambada,
which corresponds to the recommendations from the provider.

Both systems have a pay-per-query pricing model
that is based on the number of bytes in the input relations
and \SI{1}{\tebi\byte} of input costs \SI{5}[\$]{} in both systems.
Only the bytes in attributes
that are actually used in the query are taken into account
and any type of computation including complex joins are free.
However, selections are handled differently:
in \bigquery all columns are always counted in their entirety,
whereas in \athena only the selected rows of these columns are counted,
i.e., selections are ``pushed into the cost model.''
\bigquery also charges per GB-month of loaded data,
which we ignore in our comparison.

\begin{figure*}
    \centering
    \begin{subfigure}[t]{.237\linewidth}
        \centering
        \includegraphics[width=\linewidth]{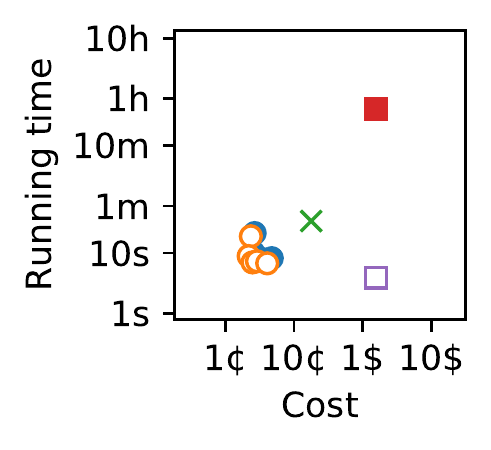}
        \caption{Q1, SF \SI{1}{\kilo\nothing}.\hspace{-4em}~}
        \label{fig:tpch-comparison-sf1k-q1-qaas}
    \end{subfigure}
    \begin{subfigure}[t]{.17\linewidth}
        \centering
        \includegraphics[width=\linewidth]{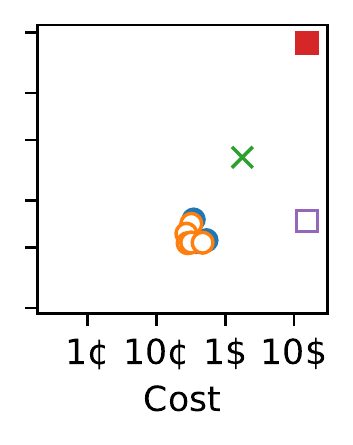}
        \caption{Q1, SF \SI{10}{\kilo\nothing}.}
        \label{fig:tpch-comparison-sf10k-q1-qaas}
    \end{subfigure}
    \begin{subfigure}[t]{.17\linewidth}
        \centering
        \includegraphics[width=\linewidth]{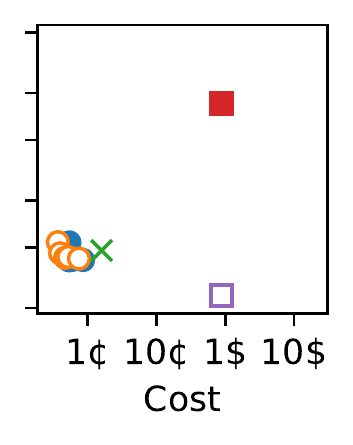}
        \caption{Q6, SF \SI{1}{\kilo\nothing}.}
        \label{fig:tpch-comparison-sf1k-q6-qaas}
    \end{subfigure}
    \begin{subfigure}[t]{.373\linewidth}
        \centering
        \includegraphics[width=\linewidth]{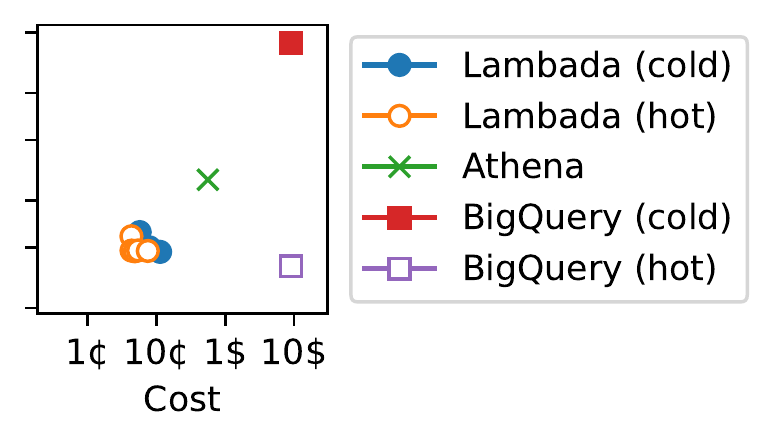}
        \caption{Q6, SF \SI{10}{\kilo\nothing}.\hspace{10em}~}
        \label{fig:tpch-comparison-sf10k-q6-qaas}
    \end{subfigure}
    \caption{Comparison of \lambada (using $F=1$ and varying $M$)
             with commercial QaaS systems.}
    \label{fig:tpch-comparison}
    \Description{Comparison of \lambada (using $F=1$ and varying $M$)
                 with commercial QaaS systems.}
\end{figure*}

We compare the three systems for TPC-H Queries~1 and 6
on scale factors \SI{1}{\kilo\nothing} and \SI{10}{\kilo\nothing}.
As described above, the latter was produced
by replicating each file of the former ten times.
We run \lambada using one worker per file ($F=1$),
i.e., using \si{320} and \si{3200} workers
for the two scale factors, respectively.
For \bigquery, we measure the time for loading the data,
add that to the running time of the query
and denote this time as ``cold'';
the query time alone is denoted ``hot.''
For \athena, we observed no noticeable difference
between the first and subsequent runs, so we only show one number.
The result is shown in Figure~\ref{fig:tpch-comparison}.

\subsubsection{Running Time}

In terms of end-to-end running time,
\lambada is the system that has the most constant latencies.
Since we use proportionally more workers as the data set grows,
the pure processing time per worker stays constant
and the latency only increases
due to the (sublinearly) larger effort for invoking the workers,
as well as a higher likelihood of stragglers and similar effects.
In contrast, \athena
does not seem to dedicate more resources for the larger data sets
since their running time increases linearly.
In BigQuery, the running time increases as well, though sublinearly,
indicating that it uses somewhat more resources for the larger scale factor.
We can only speculate why this is the case---%
at least for these simple queries,
the cloud provider could also dedicate more machines
for a shorter amount of time at an overall unchanged resource cost.
In a system like \lambada, the user has more control
and can thus increase the number of workers with the dataset size
in order to get roughly constant query latencies.

In absolute terms, compared to \athena,
the faster configurations of \lambada
are about \SI{4}{\times} faster for Q1 and on par for Q6
at SF \SI{1}{\kilo\nothing};
at SF \SI{10}{\kilo\nothing},
\lambada is about \SI{26}{\times} and \SI{15}{\times} faster, respectively.
Without taking data loading into account,
\bigquery has running times as low as \SI{3.9}{\second} and \SI{1.6}{\second}
for Q1 and Q6 at SF \SI{1}{\kilo\nothing}, respectively,
and is thus significantly faster than \lambada.
At SF \SI{10}{\kilo\nothing}, however,
it is about \SI{2.3}{\times} slower and \SI{2}{\times} faster.
Furthermore, the loading of the two scale factors takes
about \SI{40}{\minute} and \SI{6.7}{\hour}, respectively.
The loading does, hence, lead to faster querying,
but at the price of a huge delay to the answer of the \emph{first} query.
Overall, the experiment shows
that using serverless compute infrastructure
is able to provide competitive performance compared to commercial \Qaasl systems
and is even able to outperform them, in some cases by large margins.

\subsubsection{Monetary Cost}

For both queries and both scale factors,
\lambada is cheaper than both other systems.
Except for Q6 at SF \SI{1}{\kilo\nothing},
the difference is about one and two orders of magnitude
compared to \athena and \bigquery, respectively.
The difference to \bigquery is larger
even though the price per TB is the same as that of \athena
because the format of the former takes more space than that of the latter.
In these cases, the serverless approach of \lambada
is thus clearly more economic.

As expected, selections also have an influence on the cost.
While the price of Q1 is essentially the same than that of Q6 in \bigquery
(Q1 being slightly more expensive as it uses a few more attributes),
the difference is significant in \athena.
This is due to the different selectivities of the queries,
which are taken into account in \athena's pricing model.
In Q6, we only pay for the \SI{2}{\percent} of the selected rows,
while we pay for \SI{98}{\percent} of them in Q1.
For Q6, \lambada is thus only slightly cheaper than \athena.
\lambada also benefits from the selectivity
as discussed in the previous section,
but not to the same degree.
For queries with even more selective predicates,
\athena would eventually become cheaper---%
up to the point where a query becomes free
if it filters out all tuples in the input.
Even for queries where the min/max filters of Parquet work perfectly,
\lambada's cost could not be lowered below the cost
of invoking other workers, loading the plan,
fetching the metadata of each file, pruning out all row groups,
and finally returning an empty result.
In the most unfavorable case,
highly selective queries that cannot benefit from min/max filters,
\lambada would need to scan the entire input.

This discussion shows the role of the pricing model.
While a serverless query processing system like \lambada
runs on infrastructure that is rented \emph{per unit of time}
and has, thus, a monetary cost that is roughly proportional
to the amount of resources used,
the cost model of \Qaasl systems is designed
to be easily understandable by clients
and, thus, extremely simple.
It only needs to yield prices that are proportional
to the resources used by the \emph{overall workload mix}
observed by the cloud provider.
This means that some queries are necessarily under-priced
while others are over-priced,%
\footnote{As we have shown in previous work~\cite{Marroquin2018},
          it is possible to exploit this pricing model
          by executing several queries at the price of a single one.}
such as the scan-heavy queries in this section.
For this type of queries, a serverless solution like \lambada
can have the biggest advantage.

\subsection{Exchange Operator}

We compare the performance of our exchange operator
with the numbers published for similar implementations in previous work,
namely Locus~\cite{Pu2019} and Pocket~\cite{Klimovic2018b}.
We use a dataset of \SI{100}{\giga\byte}
because numbers are available for a dataset of that size
for both other systems.
Locus uses workers with \SI{1536}{\mebi\byte} of main memory;
Pocket uses \SI{3008}{\mebi\byte} workers;
for \lambada, we use \SI{2048}{\mebi\byte} of allocated memory.

\begin{table}
\caption{Running time of S3-based exchange operators.}
\label{tbl:exchange-evaluation}
\centering
\begin{tabular}{lccc}
\toprule
                & \textbf{\#Workers} & \multicolumn{2}{c}{\textbf{Storage Layer}} \\
\cmidrule{3-4}
                &                    & \textbf{VMs}     & \textbf{S3}       \\
\midrule
\textbf{Pocket}~\cite{Klimovic2018}
                & \si{250}           & \SI{58}{\second} & \SI{98}{\second}  \\
                & \si{500}           & \SI{28}{\second} &                   \\
                & \si{1000}          & \SI{18}{\second} &                   \\
\textbf{Locus}~\cite{Pu2019}
                & dynamic            &                  & \SIrange{80}{140}{\second} \\
\midrule
\textbf{Lambada}& \si{250}           &                  & \SI{22}{\second}  \\
                & \si{500}           &                  & \SI{15}{\second}  \\
                & \si{1000}          &                  & \SI{13}{\second}  \\
\bottomrule
\end{tabular}
\end{table}

Table~\ref{tbl:exchange-evaluation} shows the running time
of the various approaches.
Compared to the S3-based baseline implementation in the work on Pocket,
Lambada runs \SI{5}{\x} faster on \si{250} workers.
In contrast to that baseline, however,
Lambada's sublinear amount of requests and the usage of multiple buckets
enable it to scale to \si{500} and \si{1000} workers,
which reduces running time further.
Compared to the implementation using Pocket
(i.e., using VM-based storage for intermediate results),
Lambada is still \SI{2.5}{\x}, \SI{2}{\x}, and \SI{1.4}{\x} faster
on \si{250}, \si{500}, and \si{1000} workers, respectively.
Locus uses a dynamic number of workers
and the paper does not detail the numbers
for the experiment on \SI{100}{\gibi\byte},
but even with \si{250} workers, Lambada is about \SI{4}{\times} faster
than Locus' fastest configuration.
Compared to both other systems, Lambada has the additional advantage
of running without any always-on infrastructure.

In another experiment, we run the exchange operator
on \SI{1}{\tera\byte} and \SI{3}{\tera\byte} datasets.
It takes \SI{56}{\second} using \si{1250} workers for the former
and \SI{159}{\second} using \si{2500} workers for the latter.
On a dataset of \SI{1}{\tera\byte}, Locus takes \SI{39}{\second}
using a dynamic number of workers
(which could be higher than what we use for \lambada),
but uses VM-based fast storage for intermediate results.

\begin{figure}
    \centering
    \begin{subfigure}[t]{.75\linewidth}
        \centering
        \includegraphics[width=\linewidth]{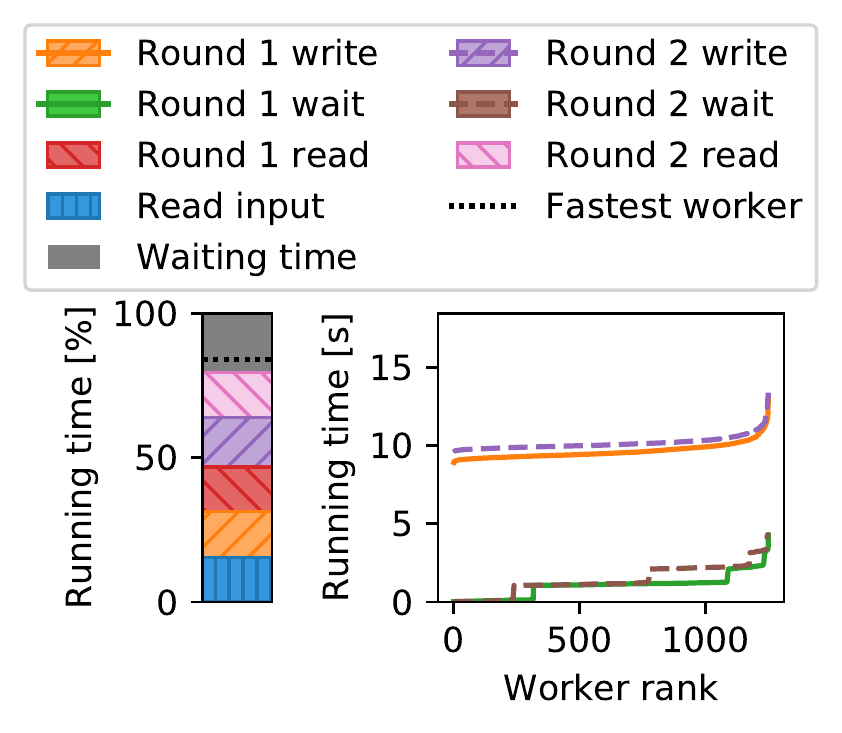}
        \caption{\SI{1}{\tera\byte}, \si{1250} workers.}
        \label{fig:exchange-latencies-1tb}
    \end{subfigure}
    \begin{subfigure}[t]{.75\linewidth}
        \centering
        \hspace{.4ex}%
        \includegraphics[width=0.914\linewidth]{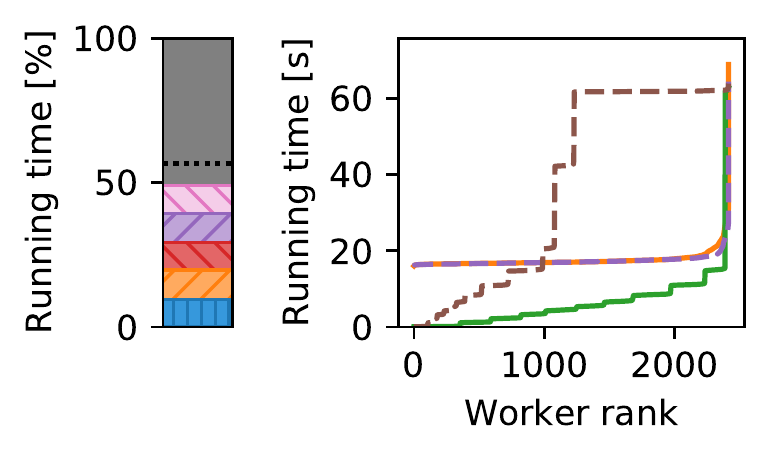}
        \caption{\SI{3}{\tera\byte}, \si{2500} workers.}
        \label{fig:exchange-latencies-3tb}
    \end{subfigure}
    \caption{Break-down and per-phase running time distribution
             of \textsc{TwoLevelExchange}.}
    \label{fig:exchange-latencies}
    \Description{Break-down and per-phase running time distribution
                 of TwoLevelExchange.}
\end{figure}

For the larger dataset (\SI{3}{\tera\byte} and \si{2500} workers),
waiting time for stragglers starts getting significant.
Figure~\ref{fig:exchange-latencies} gives details.
The left sides of the plots show the fastest running time
of each phase observed in any worker
as a fraction of the end-to-end latency
(which is dominated by the slowest worker).
Plotting the fastest execution shows an informal lower bound for each phase.
Note that reading the input,
as well as writing the partition files and reading them again
in each of the two phases,
take exactly the same amount of time
since they shuffle the same amount of data at full network bandwidth.
Also note that the fastest waiting time
is that of one round-trip to S3 (around \SI{.1}{\second}),
which is so short compared to the reading and writing
that it is not visible in the plot.
The dashed line shows the end-to-end running time of the fastest worker.
On the \SI{1}{\tera\byte} dataset,
the fastest worker takes around \SI{85}{\percent} of the slowest worker
and is relatively close to the lower bound,
i.e., to the sum of the fastest executions of the different phases.
On the \SI{3}{\tera\byte} dataset,
the total execution time is more than \SI{2}{\times} as slow
as it could be if all workers could run all phases at maximum speed;
more than half of the total execution time is due to stragglers and waiting.

The right sides of the plots give details about the stragglers.
For each phase, it gives a distribution of the running time
of each worker ordered by increasing running time.
We omit the three read phases
as they do not experience significant tail latencies.%
\footnote{This is not the case when using the default configuration.
          Instead, aggressive timeouts and retries are necessary
          to reduce tail latencies,
          but describing such optimizations in detail are out of scope of this paper.
          Then basic idea is described by Amazon's
          ``Performance Guidelines for Amazon S3''
          at \url{https://docs.aws.amazon.com/AmazonS3/latest/dev/optimizing-performance-guidelines.html\#optimizing-performance-guidelines-retry}.}
On both datasets, the write phases have a relatively stable running time
until the 95-percentile.
The slowest worker, however, is about \SI{30}{\percent} and \SI{4}{\times}
slower than the median for the small and big datasets, respectively.
These latencies propagate:
The waiting time in the first round is significant
for a large number of workers
because each worker that is slow with writing
causes wait time for all workers in its group.
In turn, those workers start later with the next phase
and thus cause wait time for even more workers.
While the wait time is moderate for the small dataset,
it dominates the execution time of the larger one.
Further research is required
to reduce the tail latencies appearing at these scales.
Nonetheless, our experiments show that exchange operators
can be implemented under a purely serverless paradigm
and even outperform approaches with always-on infrastructure.

%% file: relwork.tex
\section{Related Work}
\label{sec:rel-work}

Our work has two main lines of related work. The first one relates us with systems
using serverless infrastructure to perform distributed computations. The second
one relates to the use case for which serverless computing is a good fit, i.e.,
data analytics on cold data.

Recently, there have been many systems design proposals that leverage
\textit{serverless} functions in different settings. They range from video
analytics \cite{fouladi19,sprocket18} to data analytics \cite{flint18,Pu2019}.
For instance, Sprocket~\cite{sprocket18} uses parallel instantiations of
serverless workers (up to 1,000-way concurrency) to process a full-length HD
movie at lower costs.  Elgamal et. al.~\cite{costless18} proposed different
techniques to use serverless workers in edge computing by optimizing their cost
through function placement and fusion. We relate to these works in the sense
that we also observe that serverless workers are suitable for infrequent usage
due to their monetary costs, thus we share the motivation of optimizing our
computation to save money.

The \texttt{gg} system~\cite{fouladi19} states that the startup time of 1000
serverless workers can be done in around \SI{6}{\second}, making an improvement over
previous related work. We characterize serverless worker invocation and
determine that even if this is done this from the driver itself, it should not
take more than \SI{4}{\second} if done with concurrent spawning threads. However, in
order to achieve interactive data analytics at the scale of thousands of
workers, this is not enough.  That is why we propose using a propagation tree
strategy to invoke several thousands of workers.  This results in Lambada
managing to start several thousand workers in under \SI{4}{\second}.

\paragraph{Serverless data analysis systems.} In order to make data analytics
possible with serverless workers, there have been different attempts to achieve
a general and performant solution.  For instance, PyWren~\cite{pywren}
leverages code annotations to help users to deploy and coordinate computations
in serverless workers. 
However, parallelizing operators and other possible optimizations (e.g.,
selection and projection push-downs) still have to be done manually by the
user. In our system, such optimizations are done automatically through a series
of rewrites of our intermediate representation and then lowered into optimized
native code. Another system aiming to
perform data analytics is Flint~\cite{flint18}. The authors propose a rewrite
of the Apache Spark execution layer and use a combination of serverless workers
and storage to perform the work. However, query execution time achieve by Flint
is more than \SI{10}{\times} lower for similar queries run in Lambada. For instance, Flint
could take around \SI{100}{\second} for scanning a \SI{1}{\tera\byte} of data whereas
Lambada would take \SI{10}{\second}. This is due to a combination of Lambada using query
compilation to produce specialized data structures and tight loops for queries,
and using careful design of our scan operator.

\paragraph{Systems for storing intermediate data.} Work on systems such as
Pocket~\cite{Klimovic2018,Klimovic2018b} and Locus~\cite{Pu2019} study and
propose systems for performing shuffle operations with serverless workers.
Klimovic et al.~\cite{Klimovic2018,Klimovic2018b} focused on designing an
elastic, fast, and fully managed storage system for data with low durability
requirements. Their motivation, however, lies in a shuffle operator, which has as
main drawback the total number of requests for exchanging data among workers,
which is quadratic on the number of workers. Thus, if AWS S3 was used for exchanging
data, this would result in I/O rate limit errors when using hundreds of
serverless workers. Using the shuffle operator as motivation, Pu et
al.~\cite{Pu2019} also design a system for intermediate data that uses a
combination of AWS ElasticCache and AWS S3.  Lambada's shuffle operator is on a
different complexity class. It requires $O(\sqrt(P))$ messages to be exchanged
among $P$ serverless workers. Moreover,
\cite{Klimovic2018,Klimovic2018b,Pu2019} propose using additional services
which are not serverless by our definition, i.e., one has to provision the
total number of nodes needed in advance.

\paragraph{Critique to serverless computing.} Hellerstein et
al.~\cite{hell_lambdas} argue that serverless functions might not be suitable
for data analytics.  The authors show that for certain applications such as
machine learning workloads (doing training or prediction), using serverless
workers is not the right solution.  In this work, we show that by carefully
designing and implementing a system, serverless workers can be used for
processing cold data in an exploratory manner. Also, we show where the
driving costs are when building a data analytics engine based on purely
serverless components.

\paragraph{Data analytics on cold data.} Performing data analytics over cold data
has been largely studied from many different aspects from trying to avoid
fetching data from it \cite{arf2013} to actually processing from cold storage
devices \cite{borovica}. Moreover, properly integrating the processing of cold
data in data management systems has received attention both from
academia~\cite{Levandoski2013, Debrabant2013, eldawy2014trekking} and
industry~\cite{cold_sap, cold_ibm}. We believe that using serverless workers is
cost-effective possibility for processing cold data at interactive speed.

%% file: conclusions.tex
\section{Conclusions}
\label{sec:conclusions}

In this work, we show that serverless computing is viable and attractive
for interactive analytics on cold data.
Through the implementation of a system, \lambada,
we identify a number of challenges and propose solutions them:
tree-based invocation of workers for fast start-up,
a design for scan operators that balances cost and performance of cloud storage,
and a purely serverless exchange operator.
The latter overcomes limitations
that were previously thought to be inherent to the serverless paradigm.
Thanks to our optimizations,
\lambada can answer queries on more than \SI{1}{\tera\byte} of data
in about \SI{15}{\second},
which makes it competitive with commercial \Qaasl systems
and an order of magnitude faster than job-scoped VM infrastructure.

One interesting aspect is the difference in pricing models
compared to \qaas systems.
While the user essentially pays per \emph{accomplished work} with the latter,
a system run on top of serverless infrastructure
incurs cost for \emph{resource utilization}.
On the one hand, this means that optimizations of the system
also lowers the price for any given task that can benefit from it;
on the other hand,
bugs and difficult corner cases may be significantly more expensive.
Overall, we believe that the current price difference is large enough
to make the serverless approach attractive for a wider audience.